\documentclass[unsortedaddress,superscriptaddress,twocolumn,showpacs,preprintnumbers,amsmath,amssymb]{revtex4-1}
\usepackage{graphicx}
\usepackage{dcolumn}
\usepackage{bm}
\usepackage{latexsym}
\usepackage{color}
\usepackage{appendix}
\usepackage{tabularx}
\usepackage{ulem}
\usepackage{hyperref}
\usepackage{float}
\usepackage{tabularx}
\usepackage{color}
\usepackage{hyperref}
\usepackage{colortbl}
\usepackage{epstopdf}

\definecolor{Gray}{gray}{0.90}
\newcolumntype{a}{>{\columncolor{Gray}}c}
\definecolor{LightCyan}{rgb}{0.88,1,1}

\begin{document}

\title{Dynamics of reversals and condensates in 2D Kolmogorov flows}
\author{Pankaj Kumar Mishra}
\affiliation{Laboratoire de Physique Statistique, Ecole Normale Superieure, 24 Rue Lhomond, Paris, France}
\author{Johann Herault}
\affiliation{Laboratoire de Physique Statistique, Ecole Normale Superieure, 24 Rue Lhomond, Paris, France}
\author{Stephan Fauve}
\affiliation{Laboratoire de Physique Statistique, Ecole Normale Superieure, 24 Rue Lhomond, Paris, France}
\author{Mahendra K. Verma}
\affiliation{Department of Physics, Indian Institute of Technology, Kanpur~208 016, India}

\date{\today}

\begin{abstract}
We present direct numerical simulations of the different two-dimensional flow regimes generated by a constant spatially periodic forcing balanced by viscous dissipation and large scale drag with a dimensionless damping rate $1/Rh$. The linear response to the forcing is a $6\times6$ square array of counter-rotating vortices, which is stable when the Reynolds number $Re$ or $Rh$ are small. After identifying the sequence of bifurcations that lead to a spatially and temporally chaotic regime of the flow when $Re$ and $Rh$ are increased, we study the transitions between the different turbulent regimes observed for large $Re$  by varying $Rh$. A large scale circulation at the box size (the condensate state) is the dominant mode in the limit of vanishing large scale drag ($Rh$ large).  When $Rh$ is decreased, the condensate becomes unstable and a regime with random reversals between two large scale circulations of opposite signs is generated. It involves a bimodal probability density function of the large scale  
velocity that continuously bifurcates to a Gaussian distribution when $Rh$ is decreased further. 
\end{abstract}

\maketitle

\section{Introduction}

Two-dimensional modeling of turbulent flows has been widely used in atmospheric, oceanic, and astrophysical flows when the velocity field  depends weakly on the third dimension either due to  geometrical constraints or due to an externally applied field. 
A characteristic feature of two-dimensional turbulence is the inverse cascade of energy from the forcing scale to larger scales. In the absence of large scale dissipation, the energy accumulates in a so-called condensed flow that takes the form of a coherent vortex of the size of the domain~\cite{kraichnan:1967}.

Laboratory experiments on Kolmogorov flows, i.e. flows driven by a spatially periodic force, have been first performed to study large scale instabilities~\cite{bondarenko:1979,tabeling:1987} and nearly two-dimensional turbulence~\cite{sommeria:1986,paret:1997}. These flows are generated by applying an electromagnetic force to a thin layer of electrically conducting fluid. The linear response takes the form of a periodic array of counter-rotating vortices.  Besides the Reynolds number, $Re$, the stability of this flow  also depends on the dimensionless number $Rh$. It characterizes the linear dissipation proportional to $Rh^{-1}$, that results from the friction on the bottom boundary.  As the forcing strength is increased, various instabilities make the flow turbulent.

Stability analysis and direct numerical simulations  played a central role in getting a comprehensive picture of various dynamical features of the flow observed in the laboratory experiments.  Meshalkin and Sinai ~\cite{meshalkin:1961} performed the first linear stability analysis of the Kolmogorov flow generated by one dimensional forcing.  More complex forcings  have been investigated analytically  ~\cite{Gotoh:1984,sivashinsky:1984}.   Using direct numerical simulations, several authors have studied the sequence of bifurcations that lead to a chaotic behavior. Most of these studies consider a one-dimensional forcing.  In contrast, Braun et al.~\cite{braun:1996} studied the case of an $8\times8$ square forcing and reported several periodic branches, torus, and chaotic solutions that occur before the generation of large scale structures. Thess  ~\cite{thess:1992} was the first to take into account the friction at the bottom boundary in the stability analysis of Kolmogorov flows. He showed how the 
marginal stability curve of the linear response to the forcing depends on $Re$ and $Rh$.

In turbulent flows, friction is required to dissipate the kinetic energy transferred to large scales by the inverse cascade. When friction is too small, energy accumulates at the largest available scale. This leads to the formation of a large scale circulation (LSC)  containing most of the kinetic energy of the flow, as predicted by Kraichnan~\cite{kraichnan:1967}. When friction is increased, condensation does not occur any longer, and a turbulent flow with large scale characteristics compatible with an inverse cascade of energy is obtained. Both regimes have been observed by Sommeria~\cite{sommeria:1986} who  also reported  random reversals of the LSC~\cite{sommeria:1986}. The scaling law of the energy spectrum related to the inverse cascade has been verified more precisely~\cite{paret:1997}, and the effect of the condensate on the statistical properties of the turbulent flow has been investigated~\cite{xia:2008}. 

Tsang and Young ~\cite{tsang:2009} reported the saturation mechanism of the injected power in 2D turbulence: the large scale velocity field advects the array of vortices at the forcing scale and detunes them from the periodic forcing, thus leading to a decrease in the injected power into the flow. This process has been studied further by Gallet and Young who provided an analytical model of the condensed state in the absence of large scale dissipation~\cite{gallet:2013}. When friction is increased, LSC displays reversals. Their mean frequency increases with friction~\cite{sommeria:1986}. Reversals have been subsequently studied by Gallet et al.~\cite{gallet:2012}  using direct numerical simulations and have been compared to other reversals of a vector field such as reversals of the magnetic field generated by dynamo action. 
 
Most of the past numerical and experimental investigations focus so far on the transition from the laminar to the turbulent state of the flow. Studies that investigate the possible mechanism  leading to transitions from the condensate state to reversals are lacking. 

In this paper, we consider two-dimensional Kolmogorov flows driven by a forcing that generates $6\times6$ counter-rotating vortices. We study the parameter space and identify all the possible flow regimes together with their bifurcations. We mainly focus on the reversal and condensate regimes.

The structure of the paper is as follows. In Sec. II we present the governing equations and the numerical procedure. Sec. III discusses the different regimes of the flow obtained when the Reynolds number and the friction on the bottom boundary are varied. We present the energy budget of the flow in Sec. IV.  The dynamics of the LSC reversals is discussed in Sec V, and the condensed state is presented in Sec. VI.  We focus on the dynamical interplay between the large scale Fourier modes during reversals and in the condensate state. Finally, our conclusions are given in Sec. VII.

\section{Governing equations and numerical procedure}\label{numdetails}

We consider a numerical model of a laboratory experiment in which  a thin horizontal layer of liquid metal is submitted to a vertical magnetic field~\cite{sommeria:1986,gallet:2012}. An electric current is injected into the liquid metal through a square array of $6\times6$ alternate sources and sinks made of electrodes flush with the bottom boundary. The Lorentz force, i.e. the cross product of the radial current density in the fluid and the applied magnetic field, drives a square array of $6\times6$ counter-rotating vortices.  This flow becomes unstable and a turbulent regime can be obtained when the forcing is large compared to the dissipation. In the limit of large magnetic field, the flow is approximately two-dimensional, and it can be modeled using the following equations~\cite{sommeria:1986} written in a  dimensionless form

\begin{eqnarray}
\frac{\partial{\textbf{u}}}{\partial{t}}+ (\textbf{u}\cdot \nabla)\textbf{u}&  = &  -\nabla\sigma - \frac{1}{Rh}\textbf{u}  + \frac{1}{Re} \nabla^2 \textbf{u} + \textbf{F} ,\label{eq:NS}\\ 
\nabla \cdot \mathbf{u} & = & 0,
\label{eq:NS_incom}
\end{eqnarray}
where $\textbf{u} = (u,v)$ is the velocity field that obeys the incompressibility condition (\ref{eq:NS_incom}), and $\sigma$ is the pressure field. The second term on the right hand side represents the frictional force in the Hartman layer at the bottom of the container. The third term is the viscous force, and $\textbf{F} = (F_x,F_y)$ mimics the Lorentz force. The non-dimensional parameters are the Reynolds number, $Re=UL/\nu$, which is the ratio of the inertial term to the viscous one, and $Rh=\tau U/L$, which represents the ratio of the inertial term to the friction on the bottom boundary. Here, $U$ is a characteristic large scale velocity, $L$ is the length of the square container, and $1/\tau$ is the damping rate related to the friction.  The above equation has been made dimensionless using the length scale $L$ and the velocity scale $U$.  

We consider free-slip boundary conditions for the velocity field 
\begin{eqnarray}
u=\partial_x{v}=0~~~~ \mbox{at}~~ x=0,1; \\ 
v=\partial_y{u}=0~~~\mbox{at}~~ y=0,1.
\end{eqnarray}
To implement the above boundary conditions,  we employ the following basis functions in our spectral simulation:
\begin{eqnarray}
 u (x,y)= \sum_{l,m}4\hat{u}_{lm}\sin{\pi l x} \cos {\pi m y},\\
 v (x,y)=-\sum_{l,m}4\hat{v}_{lm}\cos{\pi l x } \sin {\pi m y},
\end{eqnarray}
where the positive integers $(l,m)$  represent  the wavenumber indices along the $x$- and $y$- directions respectively. Note that in the physical space, the indices $(l,m)$ correspond to $l$ vortices along the $x$ direction, and $m$ vortices along the $y$ direction.  Note that $\hat{u}_{lm}$ and $\hat{v}_{lm}$ are real.

The above non-dimensional equations [Eqs.~(\ref{eq:NS}-\ref{eq:NS_incom})] are solved using an object-oriented pseudo-spectral code Tarang-1.0~\cite{verma:2011}. We employ the fourth-order Runge-Kutta scheme with dynamically adjusted $dt=\Delta x/20\sqrt{E_u}$ (the CFL condition) for time advancement.  Here $\Delta x$ is the grid resolution, and $E_u$ is the total kinetic energy.  The 2/3 dealiasing scheme is used for all the runs.

The applied forcing is
\begin{eqnarray}
 \mathbf{F} & = & F_0
 \left[
 \begin{array}{c}
 \sin(6\pi x)\cos(6\pi y) \\
 -\cos(6\pi x)\sin(6\pi y) \\
 \end{array}
 \right],
 \end{eqnarray}
where $F_0=0.5$ is the amplitude of the force. The present form of the forcing is chosen to generate an array of $6\times6$  counter-rotating vortices in the limit of small $Re$ or $Rh$~\cite{sommeria:1986,gallet:2012}.  We perform runs for a wide range of parameters: $Re$ ($Re= 10^2$-$10^4$) and $Rh$ ($Rh=10^{-2}$-$10^2$).  We choose a uniform grid varying from $64^2$ to $512^2$ depending on the parameter values.  The adequacy of the resolution is verified by performing a grid-independence test.

\section{The different flow regimes} 

\begin{figure}[htbp]
\begin{center}
\includegraphics[scale=0.3]{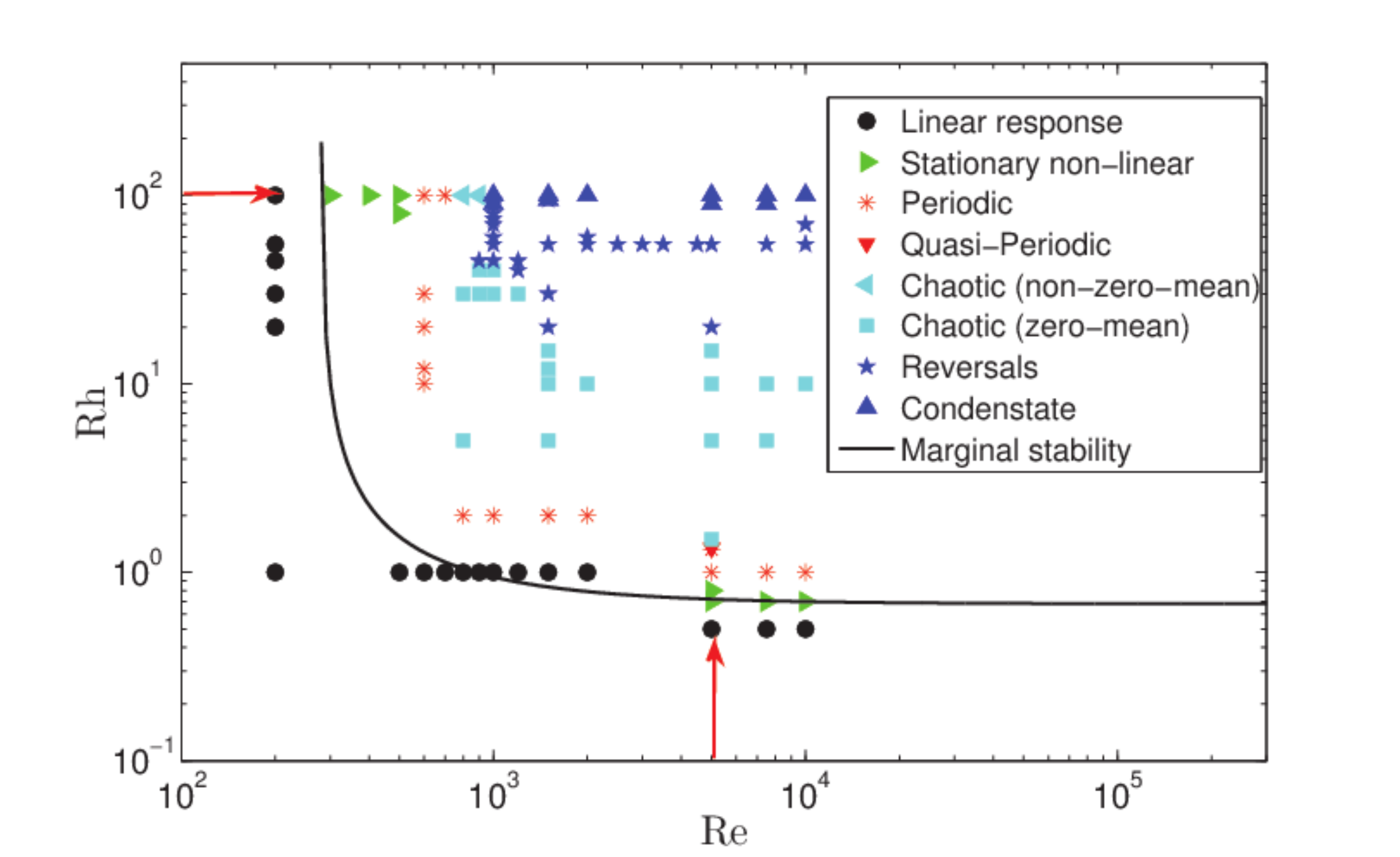}
\end{center}
\caption{The different flow regimes in the parameter space ($Re$-$Rh$):  linear response to the forcing ($\bullet$), stationary nonlinear regime ($\blacktriangleright$), time-periodic regime ($\ast$), quasi-periodic regime ($\blacktriangledown$),  chaotic regime with non zero mean flow ($\blacktriangleleft$), chaotic regime with zero mean flow ($\blacksquare$), random reversals of the large scale flow ($\bigstar$), condensate state ($\blacktriangle$). The full line is the marginal stability curve of the linear response to the forcing calculated by Thess~\cite{thess:1992}.}    
\label{fig:phase_diag}
\end{figure} 

The flow generated by the spatially periodic forcing  (for a given $F_0$ and wavenumber $k=(6\pi,6\pi)$) depends on $Re$ and $Rh$.  For low $Re$ and low $Rh$,
the  solution of Eq.~(\ref{eq:NS}) is proportional to the forcing,  i.e.,
\begin{equation}
\hat{u}_{66} \left( \frac{1}{Rh} +  \frac{72 \pi^2}{Re} \right) = F_0,  
\label{eq:laminar}
\end{equation}
which is a stable solution. These states are plotted using black filled circles in Fig.~\ref{fig:phase_diag}.  
Since Eq.~(\ref{eq:NS}) involves two dissipative terms,  the effective Reynolds number with forcing on $(6,6)$ mode can be defined as
\begin{equation}
Re_{eff} = \frac{1}{\frac{1}{Rh} +  \frac{72 \pi^2}{Re}}. 
\end{equation}
We expect the linear response to the forcing to be stable as long as $Re_{eff}$ is less than a certain constant value of order one. This justifies the form of the fit of the results obtained by Thess~\cite{thess:1992} that we have plotted as  the marginal stability curve in Fig.~\ref{fig:phase_diag}:
\begin{equation}
\frac{0.68}{Rh} +  \frac{280}{Re} = 1.
\end{equation}
Although the aforementioned fit is in a fair agreement with the data obtained by Thess, more complex behaviors exist in Thess' detailed analysis, such as bicritical points at which the geometry of the most unstable mode changes. The instability generates a flow at the largest possible scale in the limit of large $Rh$, whereas subharmonic patterns can be observed first for smaller $Rh$. It is not the purpose of our study to find these stability limits using direct numerical simulations. When we increase the parameters $Rh$ or $Re$, modes other than $(6,6)$ get generated; these new modes  are saturated by the nonlinear terms $\mathbf u \cdot \nabla \mathbf u$.  

\begin{figure}[htbp]
\begin{center}
\includegraphics[scale=0.35]{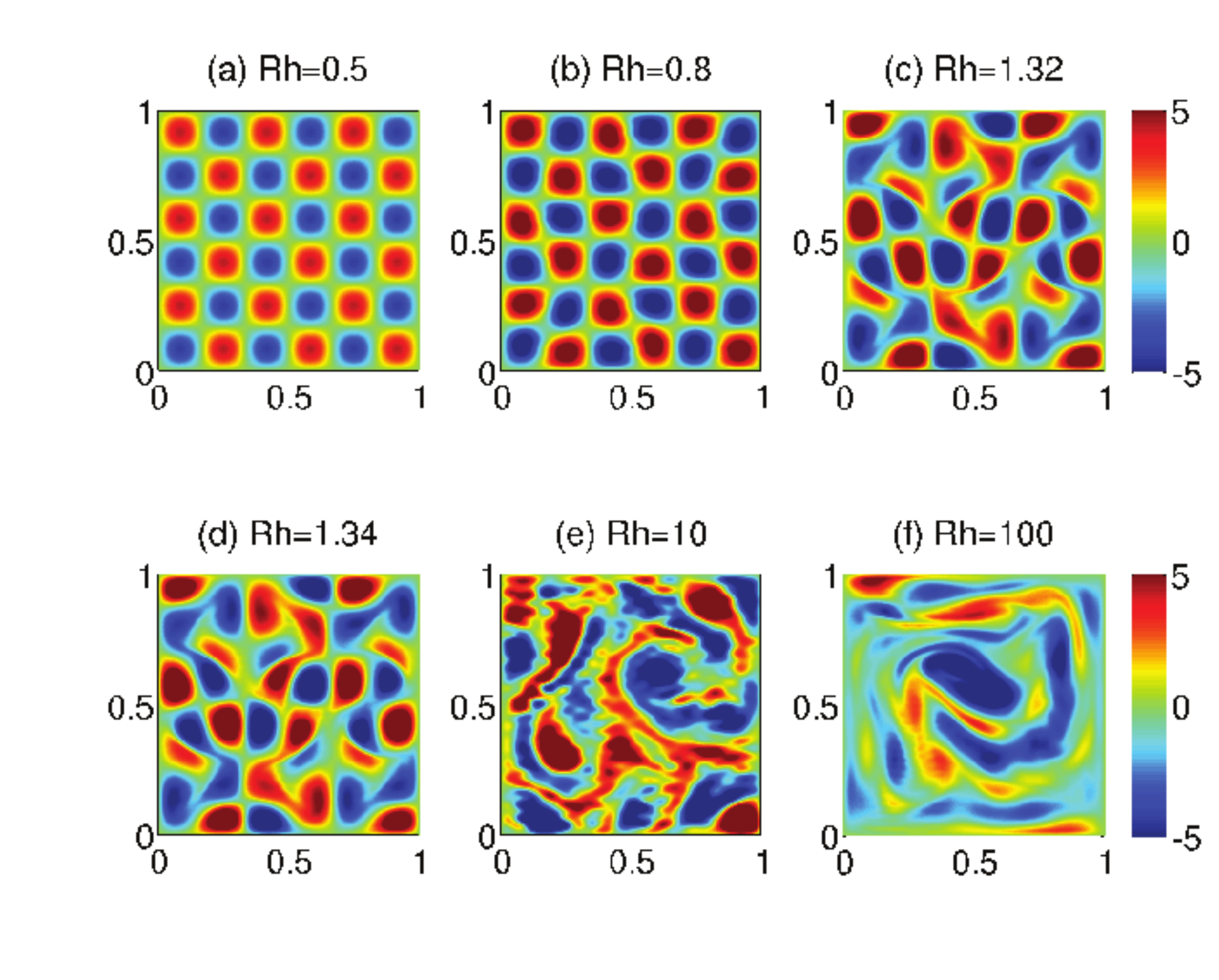}
\end{center}
\caption{Vorticity patterns for $Re=5000$: (a) Linear response to the forcing ($Rh=0.5$), (b) stationary non-linear regime ($Rh=0.8$), (c) time-periodic regime ($Rh=1.32$), (d) quasiperiodic regime ($Rh=1.34$), (e) chaotic flow with zero-mean velocity ($Rh=10$), and (f) condensate state ($Rh=100$).}    
\label{fig:phase_space_lowRh}
\end{figure}  
\subsection{Bifurcation from laminar to chaotic states}\label{bifurc}
We describe first a sequence of bifurcations from laminar to complex flows that are observed along two lines of the parameter space displayed in Fig.\ref{fig:phase_diag}, first increasing $Rh$ for $Re=5000$ (vertical arrow), then increasing $Re$ for $Rh=100$ (horizontal arrow). 

For $Re=5000$, the patterns corresponding to the linear response to the forcing are stable for $Rh\lesssim0.5$. The flow consists of a $6 \times 6$ square array of counter-rotating vortices shown in Fig~\ref{fig:phase_space_lowRh}(a).  For $Rh\sim0.8$, we observe an emergence of  modes other than the forcing mode, which is quite evident from the distortion of the flow structure (see Fig~\ref{fig:phase_space_lowRh}(b)).  At higher $Rh$, $Rh=1.32$, merger of similar signed vortices is illustrated by the snapshot displayed in Fig.~\ref{fig:phase_space_lowRh}(c). This regime involves periodic oscillations of the vortices. A further increase of $Rh$ yields successive transitions to quasiperiodic, and then to chaotic patterns (details to be discussed below); here the flow  becomes more and more disordered in space, as shown in Figs.~\ref{fig:phase_space_lowRh}(d,e). However, a different trend is observed for very large $Rh$ for which the $\hat{u}_{11}$ mode becomes dominant.  For $Rh=100$, an intense central  
vortex associated with a large scale circulation is  observed  in  Fig.~\ref{fig:phase_space_lowRh}(f).

\begin{figure}[htbp]
\begin{center}
\includegraphics[scale=0.35]{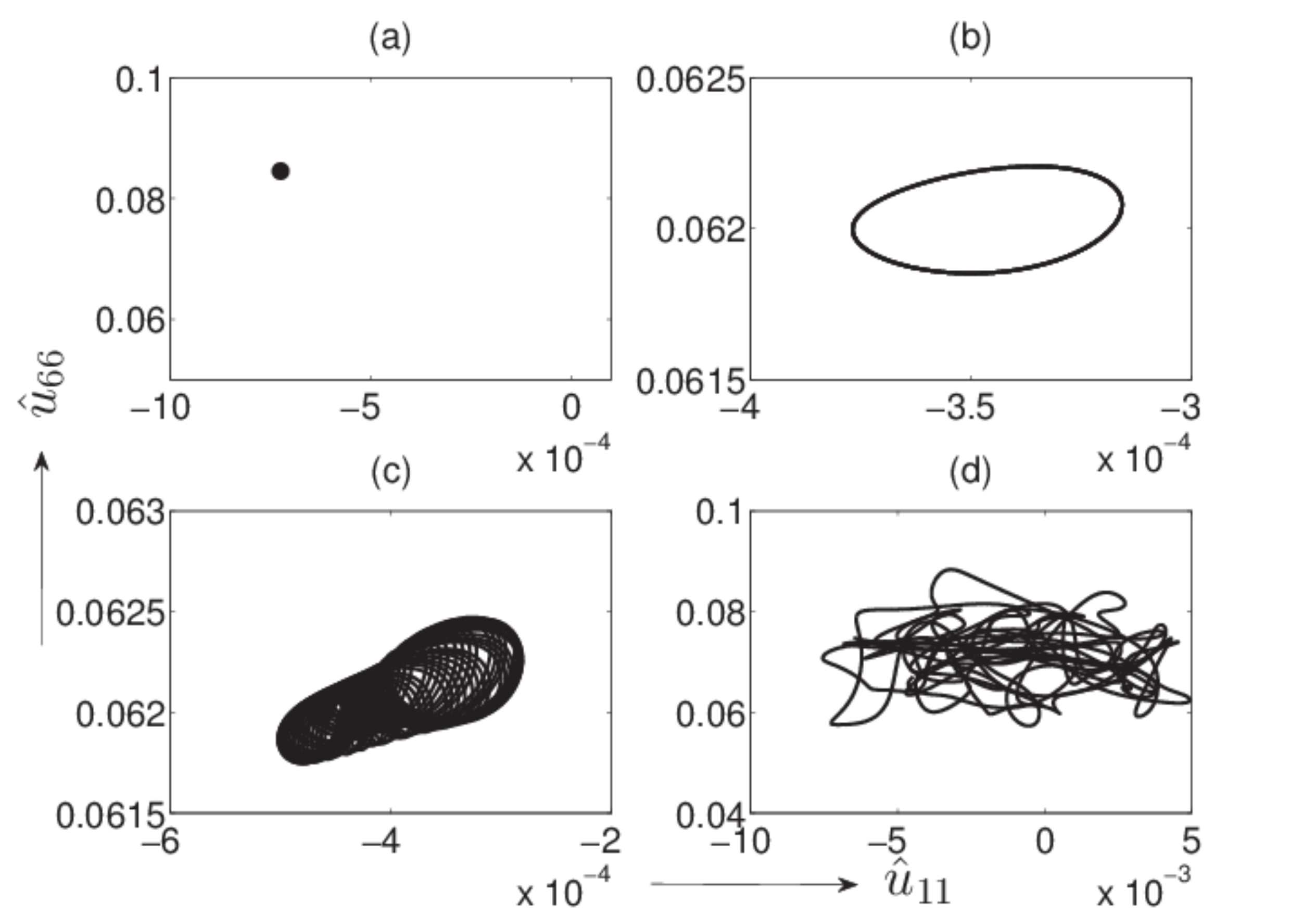}
\end{center}
\caption{Phase space projection on the forcing mode $\hat{u}_{66}$ and the large scale Fourier mode $\hat{u}_{11}$ for Re=5000 at low Rh: (a) Non-linear stationary state ($Rh=0.8$), (b) periodic regime ($Rh=1.32$), (c) Quasi-periodic regime ($Rh=1.34$) and (d) chaotic regime with zero-mean ($Rh=1.5$).}    
\label{fig:phase_space_mode}
\end{figure} 
  
We use the amplitude of the Fourier mode $\hat{u}_{11}$ to distinguish the regimes discussed above. To further quantify the dynamics,  in Fig.~\ref{fig:phase_space_mode} we plot the phase space projection along  the $\hat{u}_{11}$ and $\hat{u}_{66}$ axes. For $Re=5000$ and $Rh\sim0.8$, Fig.~\ref{fig:phase_space_mode}(a) shows a stationary state with $\hat{u}_{11} \neq 0$.   It corresponds to the pattern displayed in Fig.~\ref{fig:phase_space_lowRh}(b). These stationary states are located using right triangles in Fig.~\ref{fig:phase_diag}.  As described above, they are generated either by a  pitchfork bifurcation from the linear flow regime or by a more complex sequence of bifurcations~\cite{thess:1992}.  A further increase of $Rh$ to $1.32$ leads to a limit cycle generated through a supercritical Hopf bifurcation and displayed in Fig.~\ref{fig:phase_space_mode}(b). The corresponding spatial pattern is shown in Fig.~\ref{fig:phase_space_lowRh}(c). In Fig.~\ref{fig:phase_diag}, these limit cycles are shown 
using ($*$). For $Rh\sim1.34$, another Hopf (to be precise Neimark-Sacker) bifurcation yields a quasi-periodic state, as illustrated in Fig.~\ref{fig:phase_space_mode}(c). The spatial pattern is displayed in Fig.~\ref{fig:phase_space_lowRh}(d)). These states are located using downward triangles in Fig.~\ref{fig:phase_diag}.   This regime becomes chaotic when $Rh$ is slightly increased ($Rh=1.38$).

A further increase of $Rh$ ($Rh=1.5$) changes the system behavior to another kind of chaotic attractor, which is larger and symmetric in $\hat{u}_{11}$, as shown in Fig.~\ref{fig:phase_space_mode}(d).  For these regimes, the mean value of $\hat{u}_{11}$ vanishes, and the reflection symmetry with respect to $x=1/2$ and $y=1/2$  is statistically restored (compared to the attractor of Fig.~\ref{fig:phase_space_mode}(c)). These patterns are shown using squares in Fig.~\ref{fig:phase_diag}.  A similar sequence of bifurcations from the linear flow to a chaotic regime is observed in experiments~\cite{herault:2013}.

\begin{figure}[htbp]
\begin{center}
\includegraphics[scale=0.40]{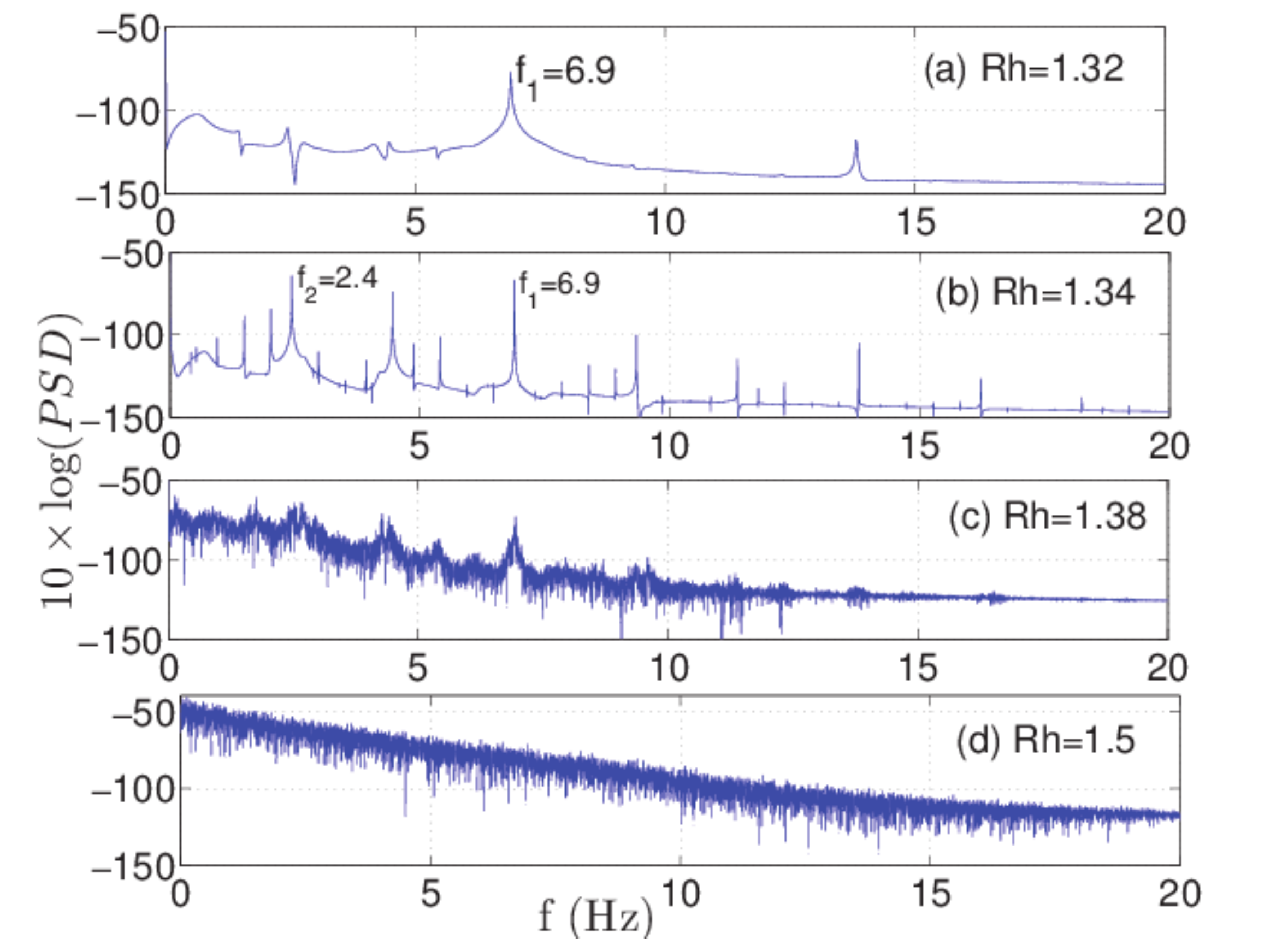}
\end{center}
\caption{Power spectral density of the mode $\hat{u}_{11}$ for $Re=5000$: (a) $Rh=1.32$ (periodic state), (b) $Rh=1.34$ (quasi-periodic state),  (c) $Rh=1.38$ (chaotic state with $\hat{u}_{11} \ne 0$) and (d) $Rh=1.5$ (chaotic state with $\hat{u}_{11}=0$).}    
\label{fig:spectrum_route_chaotic}
\end{figure} 

The power spectral densities of the large scale mode related to the periodic, quasiperiodic, and chaotic regimes described above are displayed in Fig.~\ref{fig:spectrum_route_chaotic}. The limit cycle for $Rh=1.32$ involves a fundamental frequency $f_1=6.9$ together with its harmonics (see Fig.~\ref{fig:spectrum_route_chaotic}(a)).  The power spectrum for the quasiperiodic state  corresponding to $Rh=1.34$ involves another frequency  $f_2=2.4$ and the linear combinations of $f_1$ and $f_2$ (see Fig.~\ref{fig:spectrum_route_chaotic}(b)). These peaks become enlarged by an increasing amount of low frequency noise as displayed in Fig.~\ref{fig:spectrum_route_chaotic}(c) that corresponds to a chaotic regime with a non zero mean flow ($Rh=1.38$). As $Rh$ is increased further, the two symmetric attractors merge, and a fully chaotic regime with zero mean flow is obtained (Fig.~\ref{fig:spectrum_route_chaotic}(d) for $Rh=1.5$). The noise level displays an exponential decay as a function of  frequency. 
   
\begin{figure}[htbp]
\begin{center}
\includegraphics[scale=0.45]{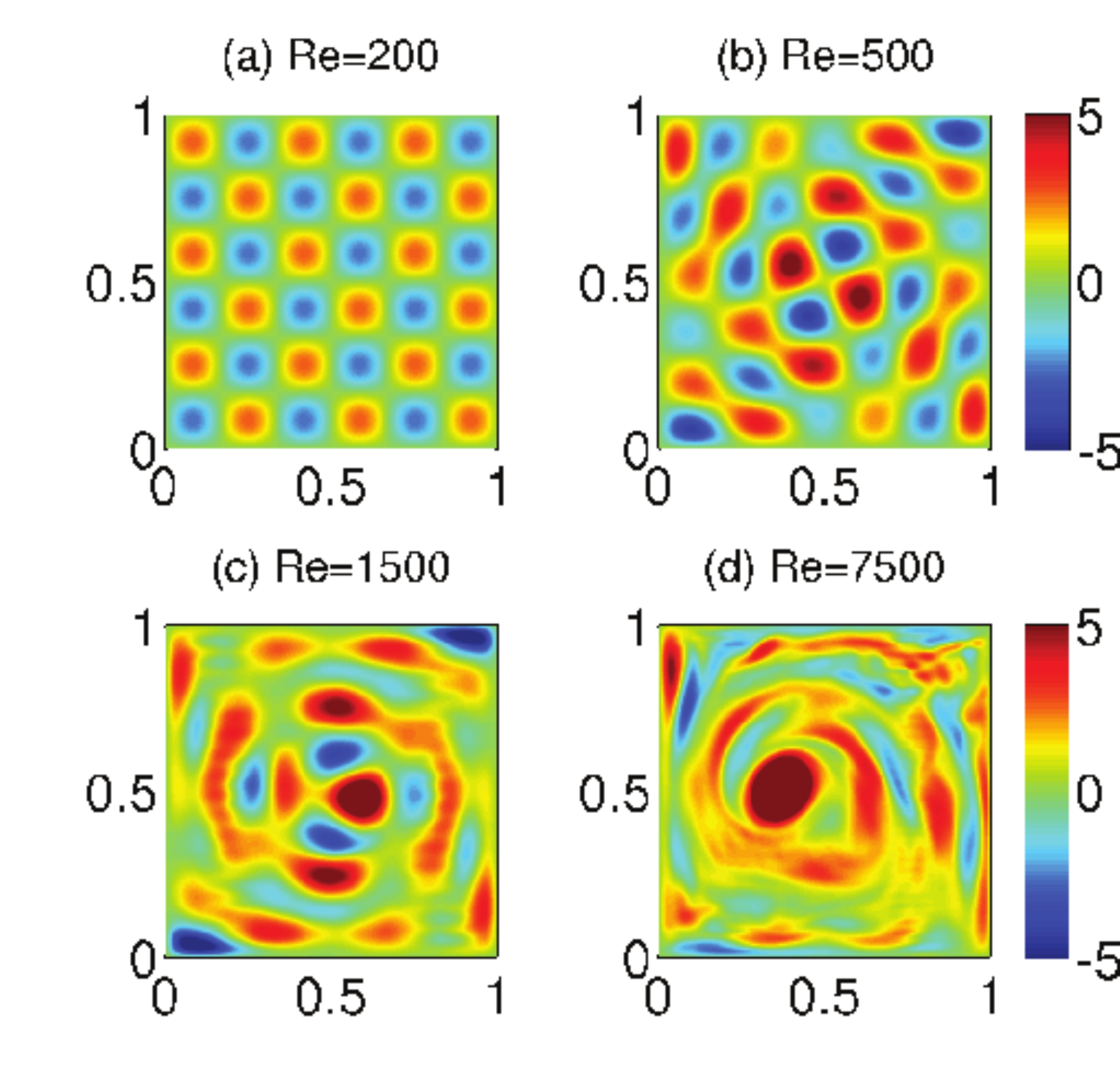}
\end{center}
\caption{Vorticity patterns for various states at $Rh=100$: (a) laminar state (Re = 200) , (b) Stationary non-linear state (Re=500),  
(c) and (d) condensate state respectively at $Re=1500$ and $Re=7500$.}    
\label{fig:phase_space_lowRe}
\end{figure}

A similar sequence of bifurcations is observed when we increase $Re$ at fixed $Rh=100$ (horizontal arrow in Fig.\ref{fig:phase_diag}).  The linear response to the spatial forcing is stable for $Re=200$ (Fig.~\ref{fig:phase_space_lowRe}(a)). When $Re$ is increased, the flow bifurcates to another stationary flow that involves  modes other than (6,6), in particular the large scale mode $(1,1)$ (see Fig.~\ref{fig:phase_space_lowRe}(b)). Further increase of $Re$ yields successive bifurcations to time-periodic and chaotic flows with non zero large scale flow. The large scale mode becomes more and more important as $Re$ is increased  further(see Fig.~\ref{fig:phase_space_lowRe}(c, d)).

In the next subsection we will describe how new patterns emerge when $Rh$ is increased beyond the chaotic regime.

\subsection{Transition from chaotic to condensate states}

\begin{figure}[htbp]
\begin{center}
\includegraphics[scale=0.45]{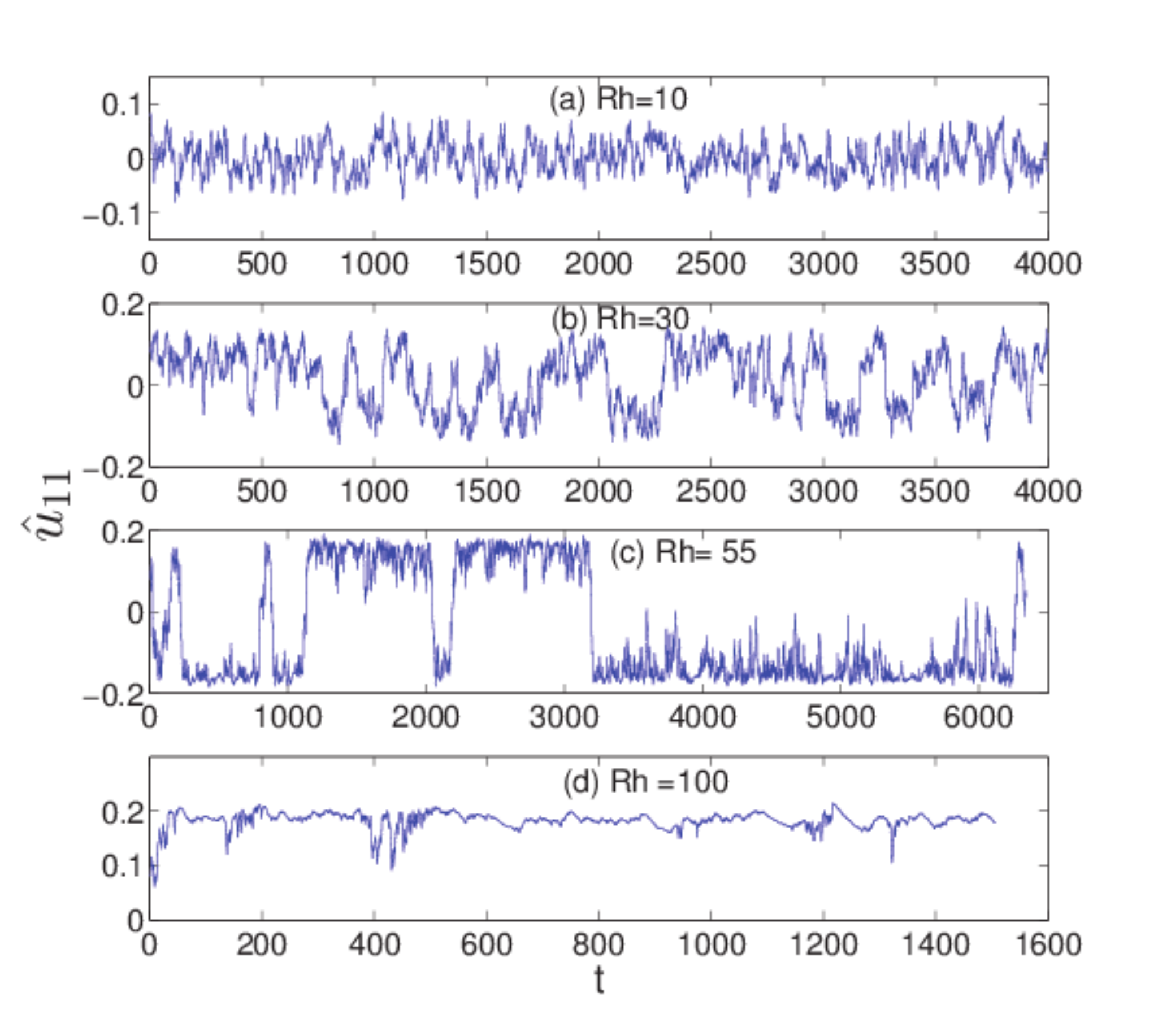}
\end{center}
\caption{For $Re=5000$: direct recordings of the large scale mode when $Rh$ is increased from $Rh=10$ (chaotic regime) to $Rh=100$ (condensed state). Random reversals between two states with opposites values of the LSC velocity are clearly  visible for $Rh=55$ but could be already guessed for $Rh=30$.}
\label{fig:direct recordings}
\end{figure}

\begin{figure}[htbp!]
\begin{center}
\includegraphics[scale=0.44]{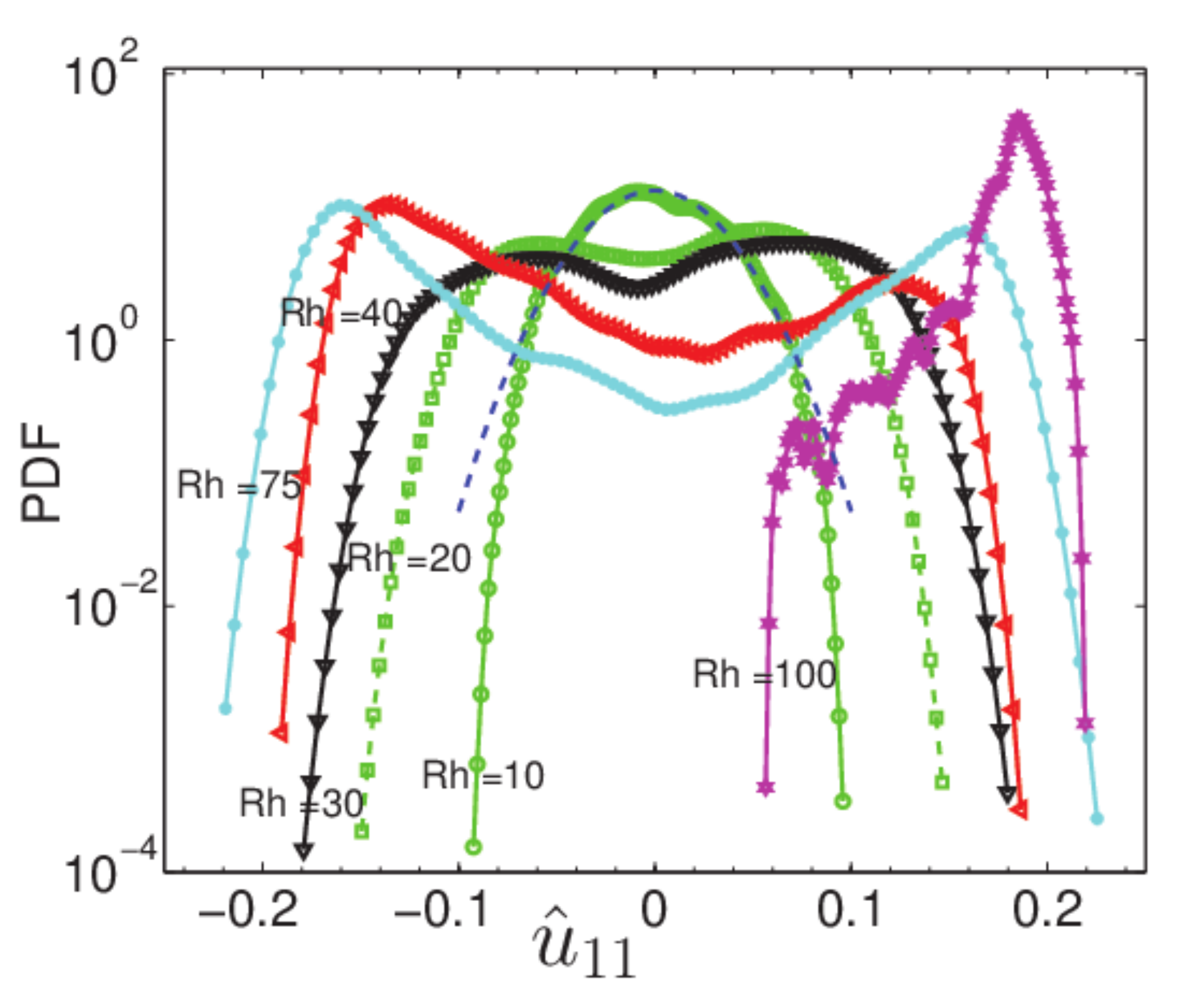}
\end{center}
\caption{Probability density function of the large scale Fourier mode $\hat{u}_{11}$ for different $Rh$ with $Re=5000$. The PDF is Gaussian for $Rh=10$ ($\circ$) and its most probable value is $\hat{u}_{11}=0$. When $Rh$ is increased, the distribution bifurcates to a bimodal PDF displayed for $Rh=20$ ($\square$). It becomes more and more pronounced as $Rh$ is increased further: $Rh=30$ ($\triangledown$),  $Rh=40$ ($\triangleleft$),  $Rh=75$ ($\bullet$).  For $Rh=100$, no reversal is observed on the simulation time. The system is in the condensed state ($\ast$). The dashed line is the Gaussian fit.}
\label{fig:PDF}
\end{figure}

In the previous subsection, we discussed in detail the system dynamics for $Rh = 0$ to 10 with $Re=5000$.  We obtained patterns corresponding to fixed point, periodic, quasiperiodic, and chaotic states.  For the chaotic state at $Rh=10$, the large scale velocity mode $\hat{u}_{11}$ randomly changes sign, as depicted in the time series of Fig.~\ref{fig:direct recordings}(a) (corresponding to Fig.~\ref{fig:phase_space_lowRh}(e)).  The probability density function (PDF) of $\hat{u}_{11}$ is displayed in  Fig.~\ref{fig:PDF}. It is close to a gaussian with a maximum for $\hat{u}_{11} = 0$.

As $Rh$ is increased further to 30, $\hat{u}_{11}$ fluctuates more strongly, as evident from the time series of Fig.~\ref{fig:direct recordings}(b).  More importantly, the PDF of $\hat{u}_{11}$ exhibits two peaks at $\pm u_0$ as shown in  Fig.~\ref{fig:PDF}.  This transition from a gaussian with a mean value $\hat{u}_{11} = 0$ to a double-peaked symmetric PDF when $Rh$ is increased can be considered as the low $Rh$ border of the reversal regime.  

A further increase of $Rh$ to 50 shows the chaotically reversing regime more clearly, as illustrated in Fig.~\ref{fig:direct recordings}(c).  The maximum of the PDF of $\hat{u}_{11}$  occurs at larger $u_0$. In addition,  the residence time in one of the states with opposite large scale circulations becomes much longer than the eddy turn-over time. The average residence time increases when $Rh$ is increased further. 
 
 Finally, when $Rh=100$, the large scale  circulation no longer changes sign, and a ``condensed" state is obtained  (see Fig.~\ref{fig:direct recordings}(d)). The large scale circulation displays intermittent fluctuations around a mean nonzero value of $\hat{u}_{11}$.   The flow pattern of a condensed state is shown in Figs.~\ref{fig:phase_space_lowRh}(f) and \ref{fig:phase_space_lowRe}(c,d).
In  Fig.~\ref{fig:PDF}, we plot the PDF of $\hat{u}_{11}$ for $Rh=100$, which shows fluctuations around a positive $\hat{u}_{11}$, consistent with the time series depicted in Fig.~\ref{fig:direct recordings}(d).  In the condensed state,  the attractors corresponding to both directions of the large scale circulation are disconnected in phase space.

The aforementioned condensed state has been attributed to the inverse cascade of energy from the forcing scale to the largest possible scale~\cite{kraichnan:1967}.  Another viewpoint has been put forward recently by Gallet and Young~\cite{gallet:2013}. They found the condensate as an approximate nonlinear solution of Eq.~(\ref{eq:NS}) in the absence of friction ($1/Rh = 0$). For the $(6, 6)$ forcing, the condensate is generated through a pitchfork bifurcation from the linear flow regime as $Re$ is increased, and it stays stationary up to $Re \sim 1000$. It is evident from our simulations that the large scale mode $(1, 1)$ starts playing a major role as soon as the linear response to the forcing becomes unstable, and this mode is  used to characterize the different flow regimes. 

Our simulations illustrate that for $Re > 1000$, the $(1, 1)$ mode is dominant  in the limit of 
large $Rh$ which corresponds to a small dissipation rate at large scales.  It is interesting to observe how the condensate disappears when $Rh$ is decreased.  We observe in Fig.~\ref{fig:phase_diag} that it bifurcates to a regime with random reversals of the large scale circulation and then to the chaotic regime with zero mean flow.

\section{Energy budget}\label{energybudget}

The aforementioned dynamical regimes involve energy exchanges among the Fourier modes due to nonlinear interactions.  These interactions generate energy transfer from the injected power by the periodic forcing to the small-scale viscous dissipation and the large-scale drag. An integration over space  of the scalar product of Eq.~(\ref{eq:NS}) with the velocity yields the following equation for the energy budget:  
\begin{equation}
 \frac{d E_u}{dt}=   I - \epsilon_{s} - \epsilon_{L}, \label{eq:ke}
\end{equation}
where 
\begin{equation}E_u = \int\int {\frac{1}{2}{\mathbf{u}^2} \,dx dy}
\end{equation}
is the kinetic energy, 
\begin{equation}
I= \int\int {{\mathbf{F} \cdot \mathbf{u}} \, dx dy} = 2F_0\hat{u}_{66} 
\end{equation}
is the injected power (note that $\hat{u}_{66} = \hat{v}_{66}$ because of incompressibility), 
\begin{equation}
\epsilon_L=\frac{1}{Rh}\int\int {{\mathbf{u}^2} \, dx dy} = \frac{2 E_u}{Rh}
\end{equation}
is the large scale dissipation, and
\begin{equation}
\epsilon_s=\frac{1}{Re}\int\int ({{\mathbf{\nabla\times u}})^2 \, dx dy}
\end{equation}
is the small scale viscous dissipation.  Note that $\epsilon_s$ is most active at small length scales, while $\epsilon_L$ is most active at large length scales since $E_u$ is strongest at large scales.  

In a statistically stationary regime, we obtain 
\begin{equation}
\langle I\rangle = \langle \epsilon_{s} + \epsilon_{L}\rangle =  \langle \epsilon_{s} \rangle + \frac{2}{Rh} \langle E_u \rangle,
\end{equation}
where $\langle \bullet \rangle$ stands for the time average.

\begin{figure}[htbp]
\begin{center}
\includegraphics[scale=0.40]{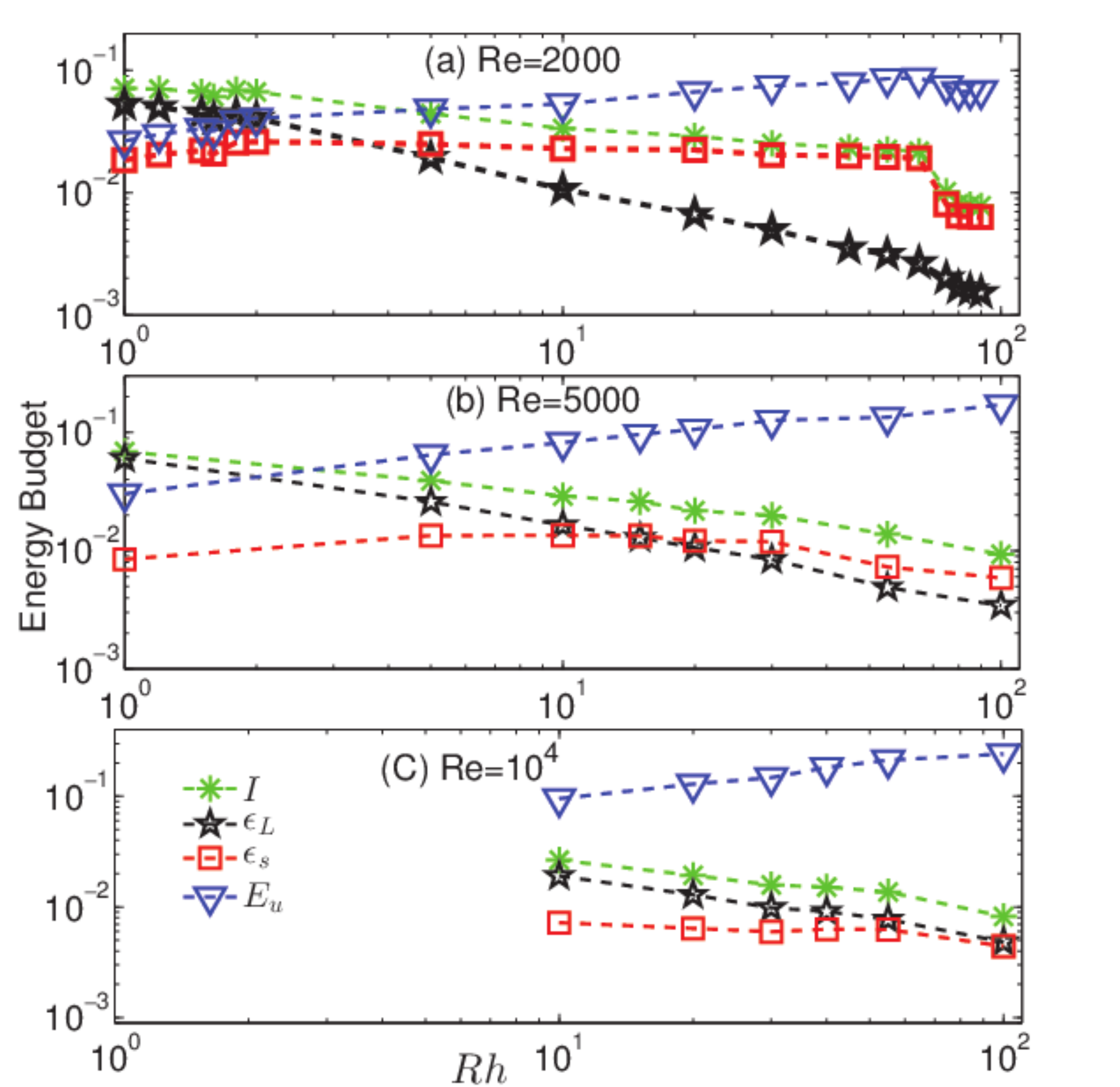}
\end{center}
\caption{Variation of the injected power, kinetic energy, and dissipation rates with Rh for: (a) $Re=2000$, (b) $Re=5000$, and (c) $Re=10^4$.}
\label{fig:energy_budget}
\end{figure}

\begin{figure}[htbp]
\begin{center}
\includegraphics[scale=0.35]{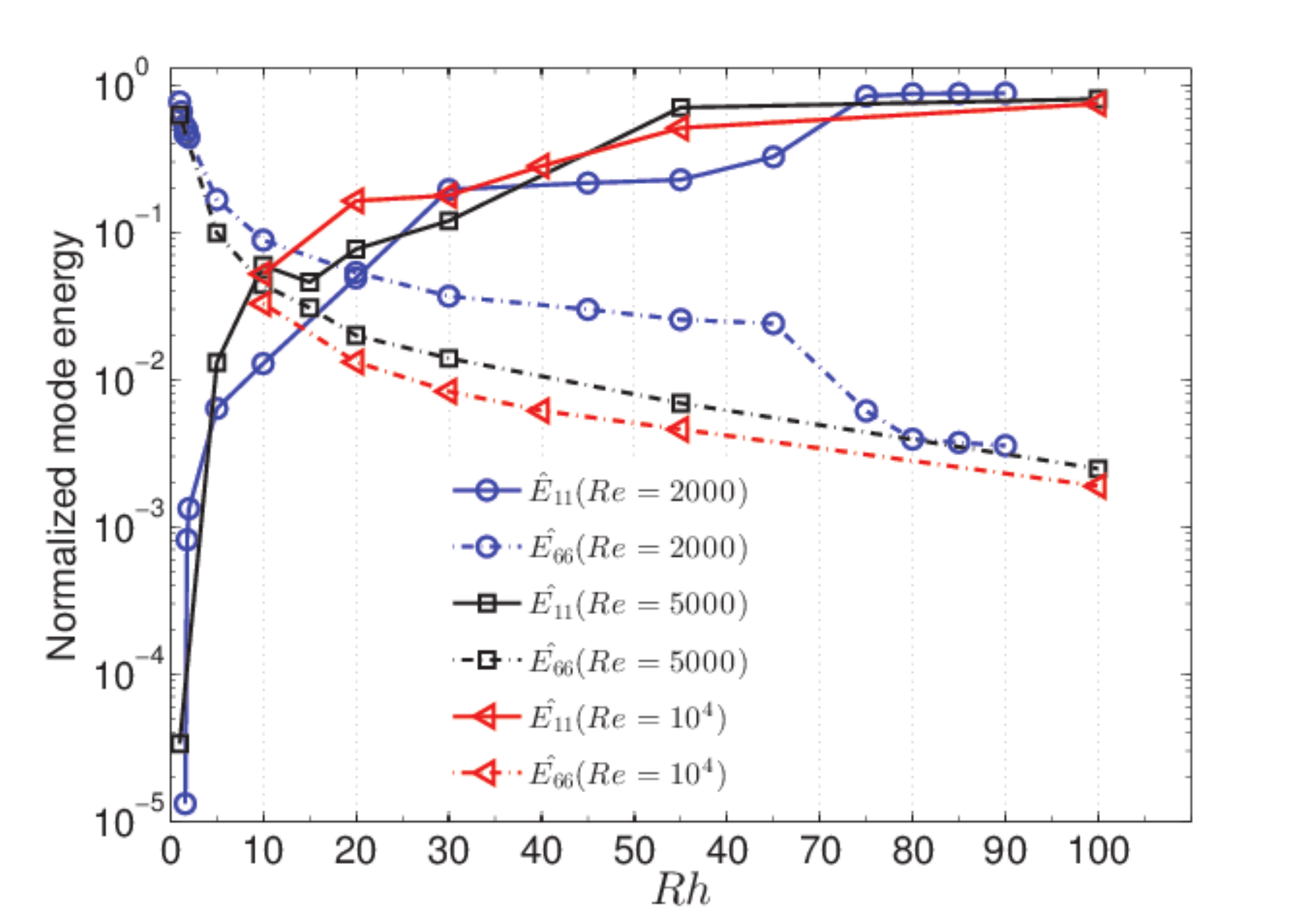}
\end{center}
\caption{Variation of the normalized energy $\hat{E}_{11}$ (respectively $\hat{E}_{66}$) of the mode $\hat{u}_{11}$ (respectively $\hat{u}_{66}$) with $Rh$ for different $Re$. }    
\label{fig:energy_ratio}
\end{figure}

In Fig.~\ref{fig:energy_budget}, we plot $I, E_u, \epsilon_L$, and $\epsilon_s$ versus $Rh$ for $Re=2000$, $Re=5000$ and $Re=10000$.   We first observe that the mean injected power $\langle I\rangle$ decreases when $Rh$ is increased. This is in agreement with the mechanism proposed by Tsang and Young~\cite{tsang:2009}; the advection of the vortices at the forcing scale by the large scale flow detunes the vortices from the spatial forcing thus decreasing its efficiency. As the energy of the large scale flow increases with $Rh$ (Fig.~\ref{fig:energy_budget} and  Fig.~\ref{fig:energy_ratio}), the detuning of the forcing becomes stronger, thus leading to the decrease of $\langle I\rangle$.

This mechanism operates here even though the present simulations are not in a regime in which viscous dissipation is negligible compared to the large scale drag. This regime was achieved in reference~\cite{tsang:2009} since the scale separation between the forcing scale and the box scale was much larger than in our simulations and hyperviscosity was used. In our case, mean viscous dissipation $\langle \epsilon_s \rangle$ and large scale drag $\langle \epsilon_L \rangle$ are of the same order in the chaotic regimes when $Re > 5000$ and $Rh > 10$. For $Re=2000$, $\langle \epsilon_s \rangle$ becomes large compared to $\langle \epsilon_L \rangle$ for the largest values of $Rh$. 

An interesting trend displayed in Fig.~\ref{fig:energy_budget} is the relative magnitude of $\langle \epsilon_s \rangle$ with respect to $\langle \epsilon_L \rangle$ when $Rh$ is varied. For $Rh=1$, the linear response to the forcing is stable and one expects $\langle \epsilon_s \rangle \ll \langle \epsilon_L \rangle$ when $Re$ is large. When $Rh$ is increased and the flow becomes chaotic, one observes that $\langle \epsilon_s \rangle$ first increases whereas $\langle \epsilon_L \rangle$ decreases. The value of $Rh$ for which they become equal increases with $Re$. 

As described above, our simulations are not performed in a turbulent regime for which $\langle \epsilon_s \rangle \ll \langle \epsilon_L \rangle$ that would correspond to the two-dimensional turbulence phenomenology in which the injected power cascades to large scales and is mostly dissipated by large scale drag. Only for the largest Reynolds number, $Re=10^4$, we have $\langle \epsilon_L \rangle$ a few times larger than $\langle \epsilon_s \rangle$ for $Rh=10$ but their ratio is of order one for $Rh=100$. We expect that $\langle \epsilon_L \rangle/ \langle \epsilon_s \rangle$ will be larger for $Re = 10^5$ which is the value of the Reynolds number reached in the experiments~\cite{sommeria:1986,herault:2013}. Increasing scale separation between the forcing and the box size will also allow the development of the inverse cascade, thus increasing $\langle \epsilon_L \rangle/ \langle \epsilon_s \rangle$. However, even though our simulations are not conducted in this regime, the  condensed state predicted by two-
dimensional turbulence is well observed.

The normalized energy $\hat{E}_{11}$ (respectively $\hat{E}_{66}$) of the mode $\hat{u}_{11}$ (respectively $\hat{u}_{66}$) is displayed  versus $Rh$ for $Re=2000$, $5000$ and $10^4$ in Fig.~\ref{fig:energy_ratio}. We define $\hat{E}_{11} = \hat{u}_{11}^2/E_u$ and $\hat{E}_{66}=\hat{u}_{66}^2/E_u$ where $E_u$ is the kinetic energy of the flow. $\hat{E}_{66}$ strongly decreases when $Rh$ is increased such that the linear response to the forcing becomes unstable, while $ \hat{E}_{11}$ increases. As said above, this phenomenon is observed even though we are not in a regime with a well developed inverse cascade with $\langle \epsilon_s \rangle \ll \langle \epsilon_L \rangle$.

\begin{figure}[htbp]
\begin{center}
\includegraphics[scale=0.35]{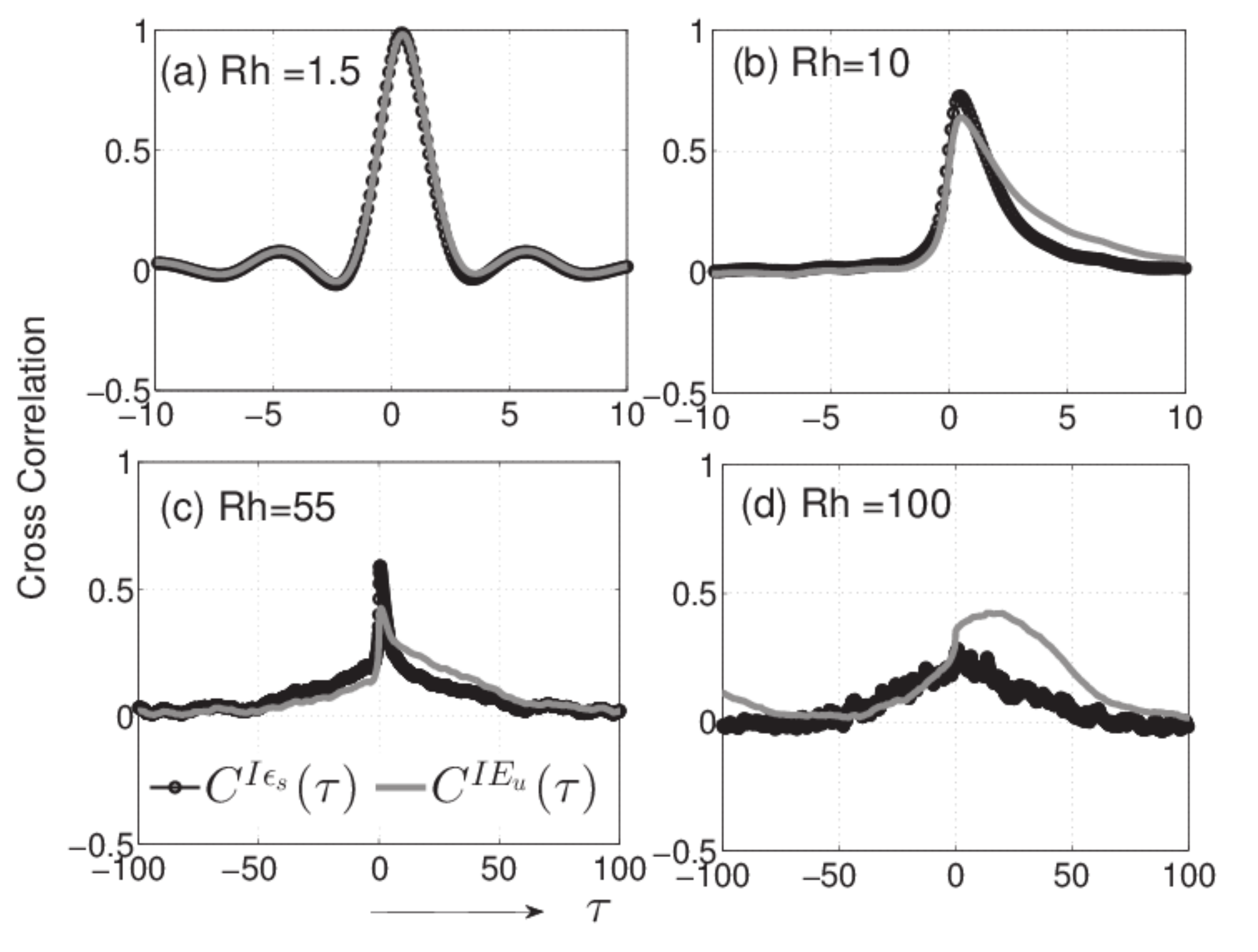}
\end{center}
\caption{Cross-correlation between the injected power $I$, kinetic energy $E_u$ and viscous dissipation $\epsilon_s$ for different values of $Rh$ and $Re=5000$.}   
\label{fig:correlation}
\end{figure}

The temporal cross-correlation between the injected power $I$ and the kinetic energy $E_u$, $C^{I E_u}(\tau)$, and the temporal cross-correlation between $I$ and  the viscous dissipation $\epsilon_s$,  $C^{I \epsilon_s}(\tau)$, are displayed in Fig.~\ref{fig:correlation}. The correlation reach unity in the chaotic regime with zero mean ($Rh=1.5$, see Fig.~\ref{fig:correlation}(a)). The maximum of the correlation decreases as $Rh$ is increased, because  the flow is much more disordered in space for $Rh=10$ or $Rh=55$  (reversal regime) than for $Rh=1.5$. For these values of $Rh$, the cross-correlation of injected power with small scale dissipation is larger than the correlation of injected power with large scale dissipation since $\epsilon_L = E_u /Rh$  (Fig.~\ref{fig:correlation}(b, c)).   In the condensed state, the opposite is observed (Fig.~\ref{fig:correlation}(d)). The maximum of the cross-correlation is for positive time lag $\tau >0$, thus kinetic energy or large scale dissipation and viscous 
dissipation lag behind injected power. In the turbulent regimes ($Rh=10$ or $Rh=55$), the time lag is of the order  of one eddy turn-over time.  

\section{Reversals of the large-scale circulation (LSC)}\label{reversal}

As described in earlier sections, random reversals of the LSC are observed for $Re > 1000$ and intermediate values of $Rh$ in a rather extended domain of  ($Re$-$Rh$) parameter space (Fig.~\ref{fig:phase_diag}).  These reversals regimes are illustrated in Fig.~\ref{fig:direct recordings}(c) using the  time series of $\hat{u}_{11}$ and plotted in Fig.~\ref{fig:phase_diag} using ($\bigstar$).   The PDF of $\hat{u}_{11}$ displayed in Fig.~\ref{fig:PDF} shows two maxima at $\pm u_0$ for the reversal states.   

\begin{figure}[htbp]
\begin{center}
\includegraphics[scale=0.35]{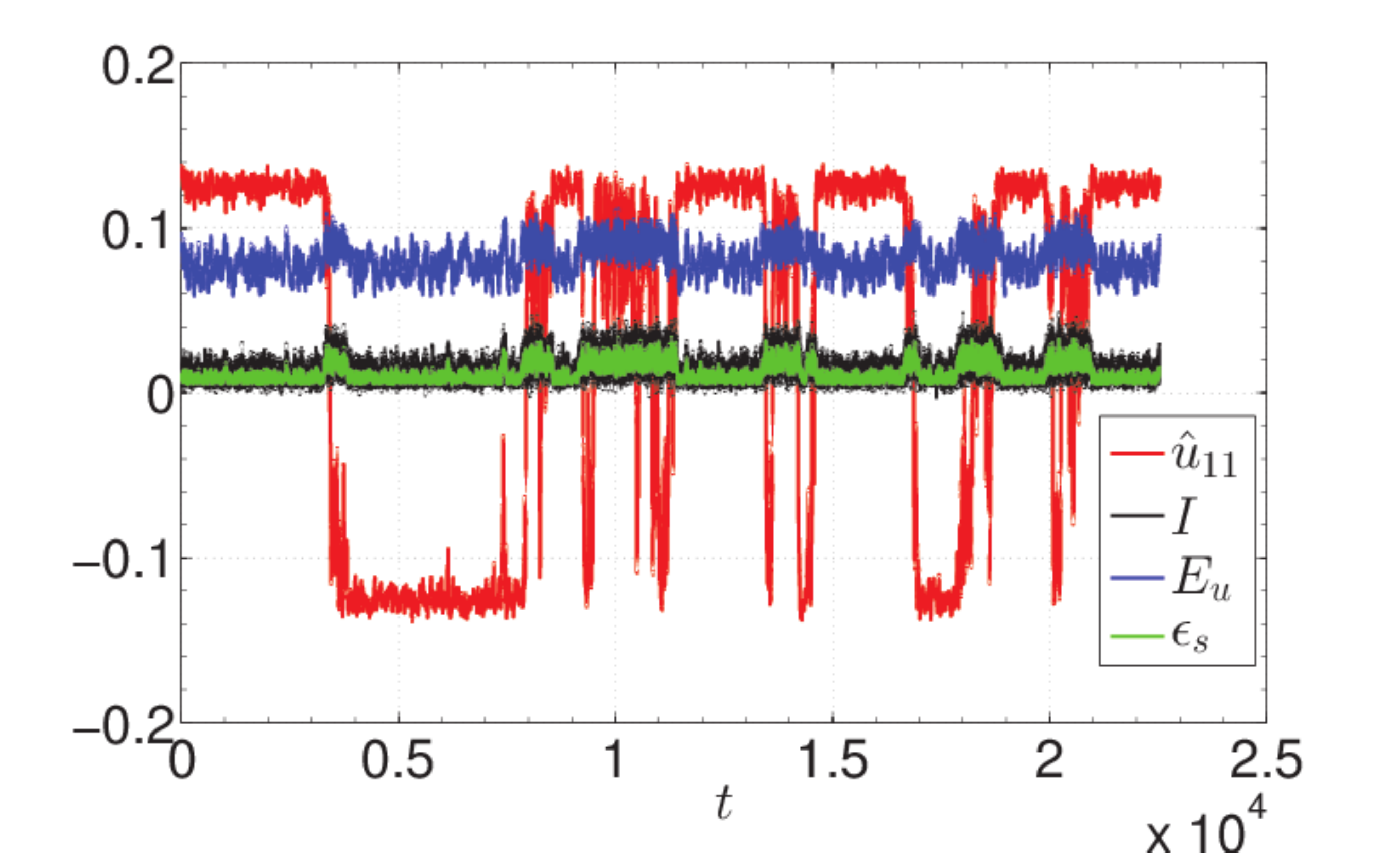}
\end{center}
\caption{Time series of  large scale circulation $\hat{u}_{11}$ (red) displaying random reversals as well as excursions for $Re=2000$ and $Rh=65$. The kinetic energy $E_u$ (blue) jumps between two states of low, respectively high, mean values. Correspondingly the injected power $I$ (black) and viscous dissipation $\epsilon_s$ (green) display weak, respectively strong fluctuations.}   
\label{fig:rev_timeseries1}
\end{figure}

Fig.~\ref{fig:rev_timeseries1} shows  time series of  $\hat{u}_{11}$,  injected power $I$, kinetic energy $E_u$, and viscous dissipation $\epsilon_s$ for $Re=2000$ and $Rh=65$.  Besides reversals, one also observes excursions or aborted reversals for which the amplitude of $\hat{u}_{11}$ starts decreasing as if a reversal were initiated, but the flow then comes back to the state with the same sign of $\hat{u}_{11}$ instead of reversing. The value of $E_u$ increases somewhat during the reversals.  The high energy state (during the transition) is  associated with  a larger dissipation at large scale, and  is also related to higher levels of injected power $I$ and viscous dissipation $\epsilon_s$.  Thus, $I$, $E_u$, $\epsilon_s$, and $\epsilon_L$ exhibit small spikes during reversals and excursions  (see below). 

In Fig.~\ref{fig:phase_space_2e03}(a,b) we present the phase-space projection $\hat{u}_{11}$-$\hat{u}_{66}$ for $(Re=2000,  Rh=65)$ and $(Re=10^4,  Rh=55)$. For $(Re=2000,  Rh=65)$,  the low energy state corresponds to the dense lobes at the bottom of the attractor. The high energy state, which also corresponds to a larger injected power, i.e. a larger mean value of $\hat{u}_{66}$, corresponds the regions with medium density. Only these regions are connected by trajectories corresponding to reversals. For $Re=2000$ and $Rh=80$, the two dense lobes get disconnected and  become the condensed state.   The presence of low and high energy injection states, corresponding respectively to high and low amplitudes of $u_{11}$, is in good agreement with the saturation mechanism mentioned above. 

For $(Re=10^4,  Rh=55)$, the low energy state no longer exists (see Fig.~\ref{fig:phase_space_2e03}(b)).  The mode $\hat{u}_{11}$ reverses between positive and negative values, but  the peak of the PDF $u_0$ is lower than that for  $(Re=2000,  Rh=65)$.  Similar features involving two turbulent states with different level of fluctuations in relation with condensation have been described in experiments~\cite{shats:2005}.

\begin{figure}[htbp]
\begin{center}
\includegraphics[scale=0.35]{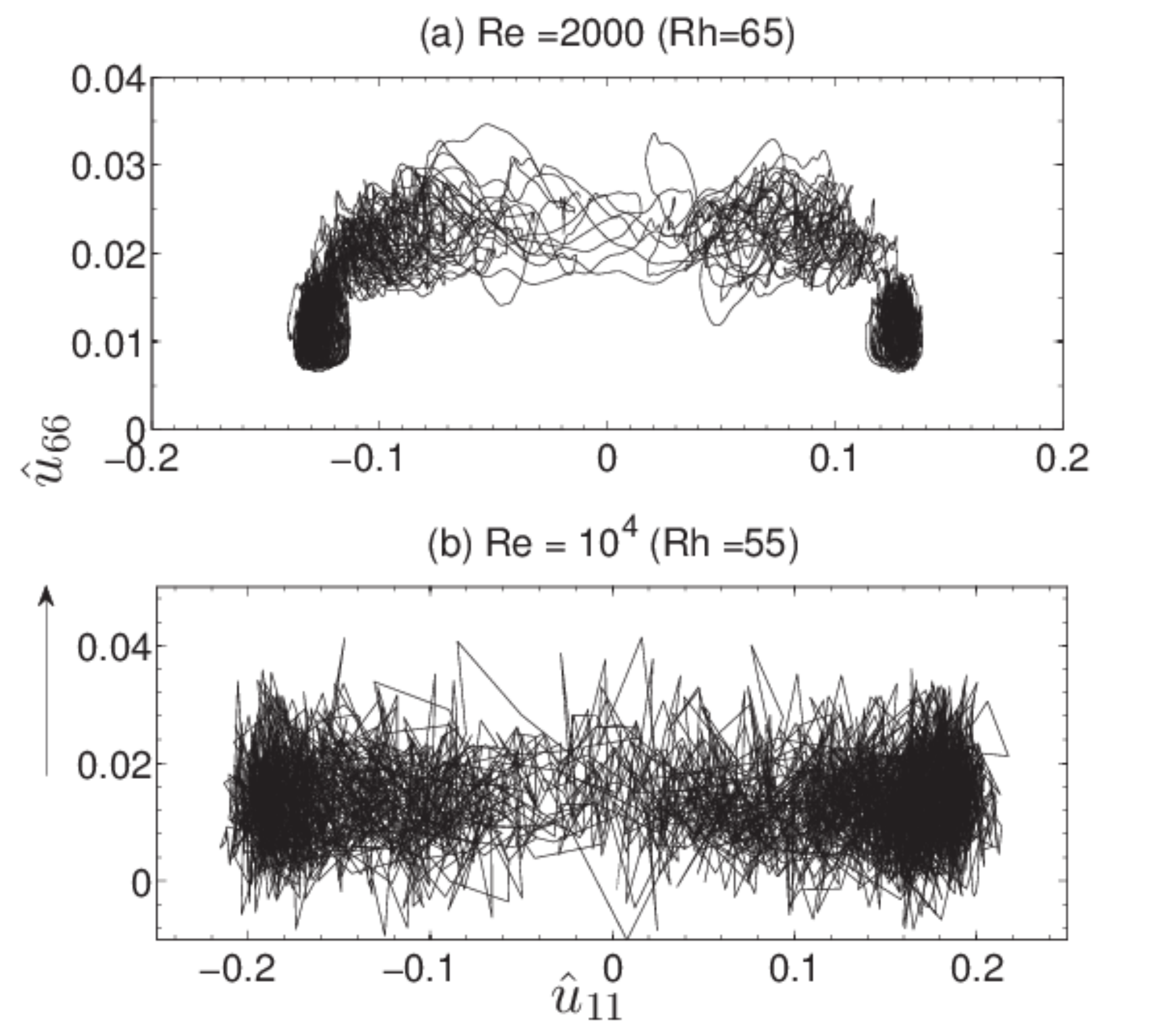}
\end{center}
\caption{Phase space projection on the modes $\hat{u}_{11}$-$\hat{u}_{66}$ for the reversal regime with $Re=2000$ (a) and $Re=10^4$ (b). Two types of attractors are displayed: one with low fluctuations and another with high fluctuations coexist for $Re=2000$, whereas the low fluctuating state is absent for $Re=10^4$. The reversals always take place between the two high fluctuating attractors.}   
\label{fig:phase_space_2e03}
\end{figure}

\begin{figure}[htbp]
\begin{center}
\includegraphics[scale=0.40]{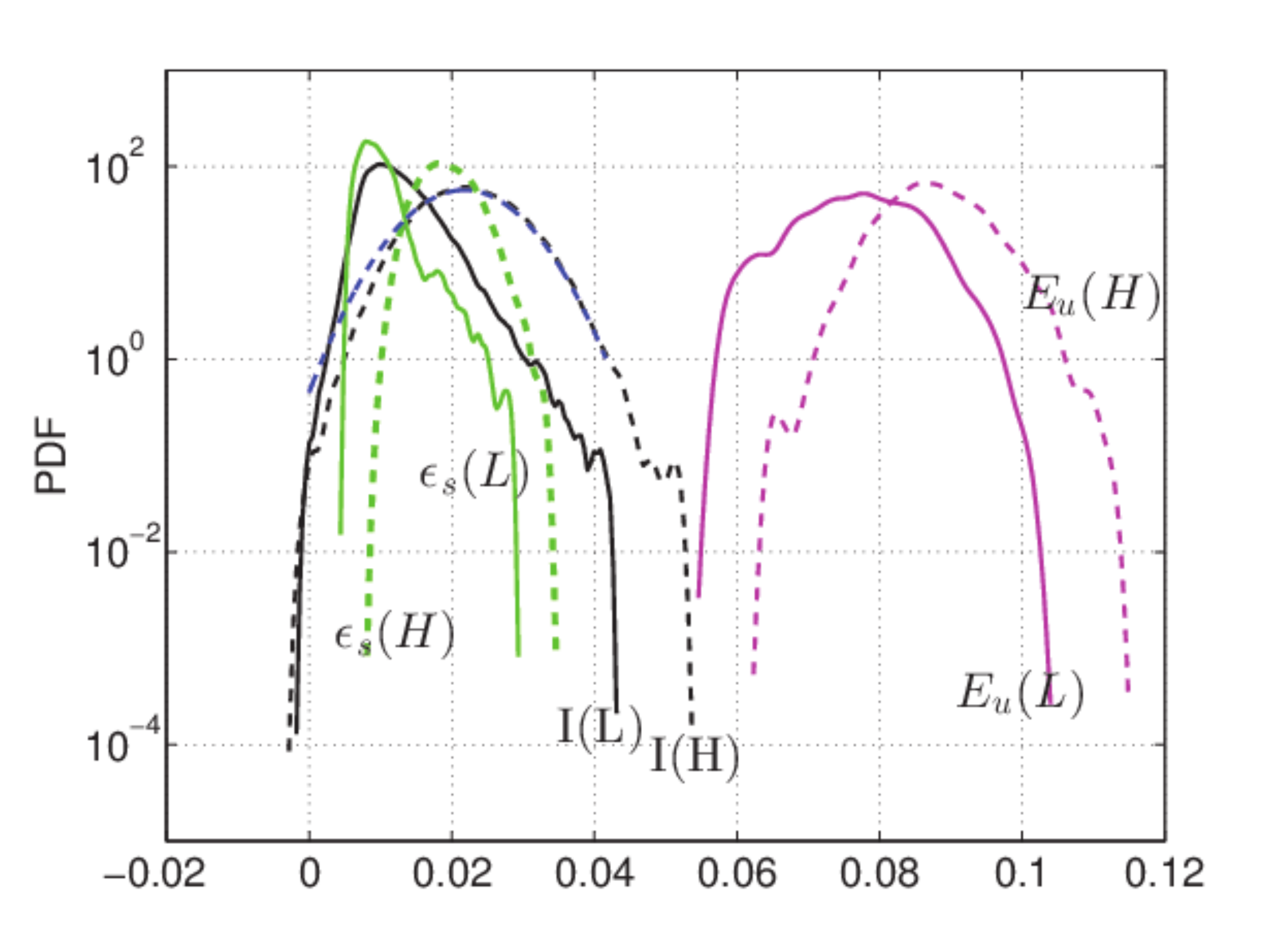}
\end{center}
\caption{PDF of the injected power (I), viscous dissipation ($\epsilon_s$)  and total kinetic energy ($E_u$) during low ($L$) and high ($H$) energy states as observed in Fig.~\ref{fig:rev_timeseries1}. The dashed line superimposed on $I(H)$ is a Gaussian fit.}   
\label{fig:PDF_Re2e03}
\end{figure}

Further information about the statistical properties of the low and high energy states can be obtained from the PDF displayed in Fig.~\ref{fig:PDF_Re2e03}.  The PDFs of $E_u$ show symmetric fluctuations about the mean in both states. In contrast, the PDF of the injected power $I$ is roughly Gaussian in the high energy state whereas it is asymmetric with an exponential tail toward large values in the low energy state. A similar trend is observed for the PDF of viscous dissipation $\epsilon_s$.  The PDFs of the low energy state have a similar shape to the ones of the condensed state.

\begin{figure}[htbp]
\begin{center}
\includegraphics[scale=0.40]{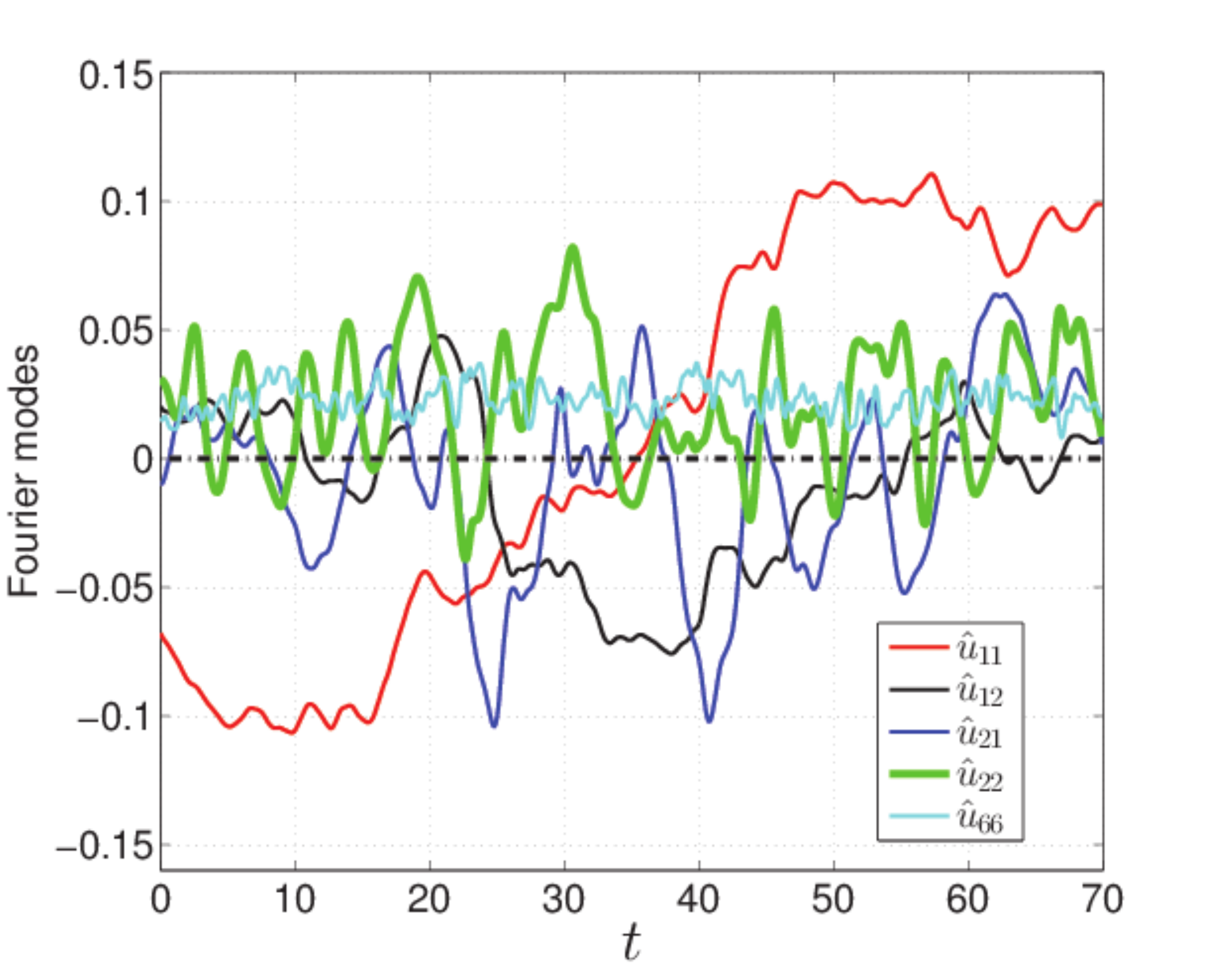}
\end{center}
\caption{Time series of the Fourier modes: $\hat{u}_{11}$ (red), $\hat{u}_{12}$ (black), $\hat{u}_{21}$ (blue), $\hat{u}_{22}$ (green), and $\hat{u}_{66}$ (cyan), for $Re=2000$ and $Rh=60$.}   
\label{fig:rev_timeseries}
\end{figure}

Next we consider the time series of some of the dominant modes during a reversal of the LSC in Fig.~\ref{fig:rev_timeseries}. We plot the time series of $\hat{u}_{11}$, $\hat{u}_{22}$, $\hat{u}_{21}$, and $\hat{u}_{66}$.  Clearly, $(1,1)$ mode switches sign, while the mode $(6,6)$ does not change sign.  The other modes display large fluctuations compared to their mean value. However, we observe that the mode $(2,2)$ keeps a positive mean. This trend is not observed for the modes $(1,2)$ and $(2,1)$ which exhibit their strongest fluctuations during the reversal. 

In  Appendix B we classify the reversing modes based on group theoretic arguments, and show that the Fourier modes $\{ E \} =$ (even, even), $\{ O \} =$ (odd, odd), $\{ M_1 \} =$ (even, odd),  $\{ M_2 \} =$ (odd, even) should behave during reversals according to a small number of possible scenarios.   One of them is: $\{ O \} \rightarrow \{ -O \}$; $\{ E \} \rightarrow \{ E \}$; $\{ M_1, M_2 \} = \epsilon$.  That is, the even modes do not change sign, the odd modes change sign, and the mixed modes $\{ M_1 \} $ and $\{ M_2 \}$ have zero mean.   A similar feature has been observed for the flow reversals in two-dimensional turbulent convection: all the odd modes (e.g., $(1,1), (3,3)$, etc.) change sign, while the even modes (e.g., $(2,2)$) keep the same sign~\cite{chandra:2011,chandra:2013}. However, in the present study, these possible scenarios are blurred by strong turbulent fluctuations. 
 
  Gallet et al.~\cite{gallet:2012} studied the evolution of $(1,1)$, $(2,1)$, and $(1,2)$ modes using numerical simulations, and obtained similar results, that is, $\hat{u}_{11}$ reverses, but  $\hat{u}_{21}$  and $\hat{u}_{12}$   fluctuate around zero, with strong fluctuations during a reversal.    Gallet et al.~\cite{gallet:2012}  based their arguments on reflection symmetries about $x$ and $y$ axes passing through the centre of the box.  The symmetry arguments of Gallet et al.~\cite{gallet:2012} and those given in the appendix are equivalent.

Gallet et al.~\cite{gallet:2012}  proposed a low dimensional model of reversals of the LSC driven by a $(6,6)$ forcing using the modes (1,2) and (2,1) which have a quadrupolar like symmetry interacting with the large scale mode (1,1) which has a dipolar like symmetry. The possible triad through which (1,2), (2,1) and (1,1) could interact is [(1,2), (2,1), (1,1)]. However, this interaction is forbidden due to the symmetry of (1,1) mode under a rotation by $\pi/2$. A closer inspection shows that the  interacting triads responsible for reversals are $\{(1,1), (2,2), (3,1) \}$ and $\{(1,1), (2,2), (1,3) \}$. 

\begin{figure}[htbp]
\begin{center}
\includegraphics[scale=0.3]{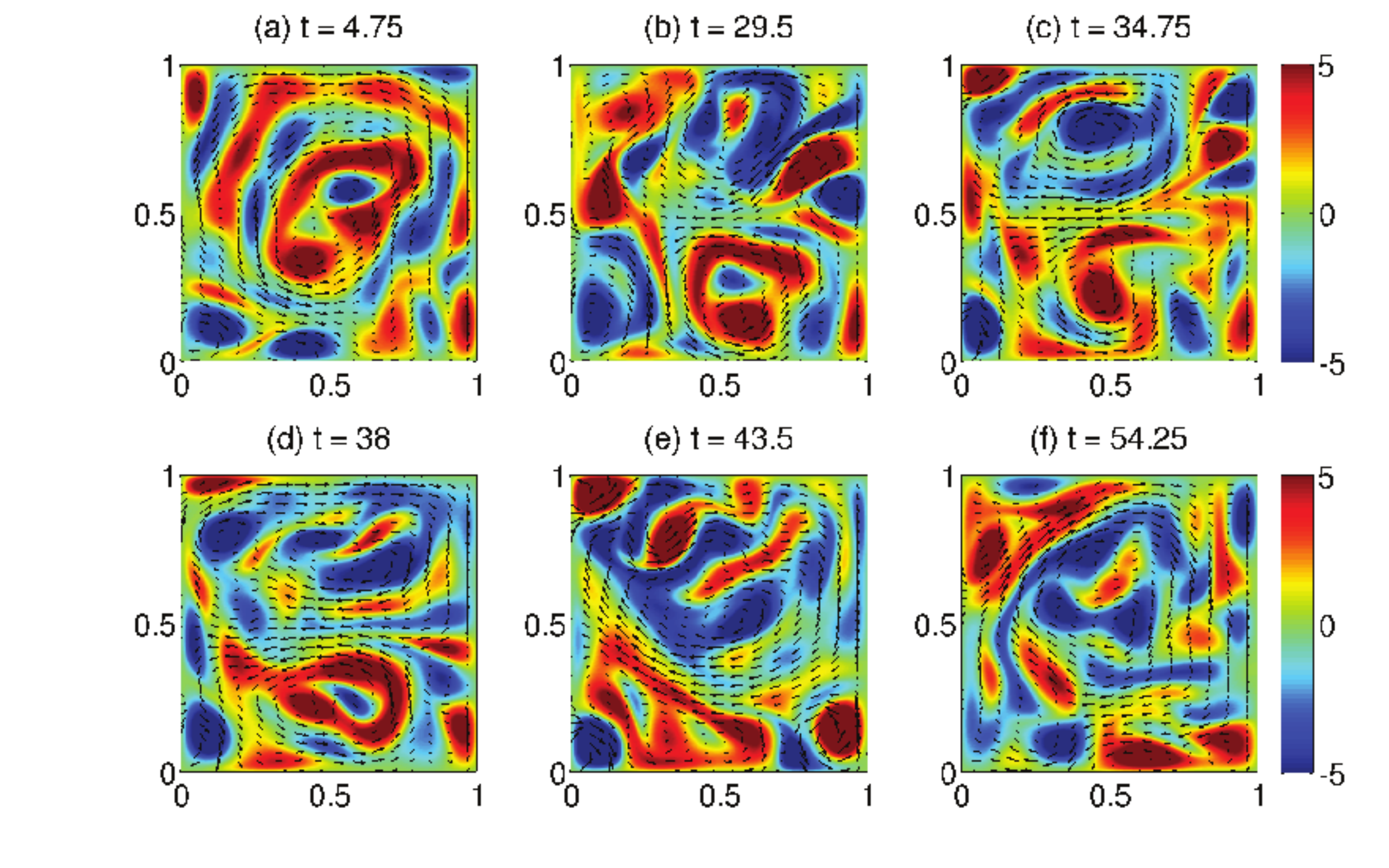}
\end{center}
\caption{Snapshots of vorticity and velocity fields during a reversal as shown in Fig.~\ref{fig:rev_timeseries}. (a) before the reversal; (b) initiation of the reversal; (c) and (d) during the reversal; (e) ending of reversal; and (f) after the reversal. (1,1) is the dominant Fourier mode before and after the reversal. (2,2) mode is active at the beginning of the reversal, followed by (2,1) and (1,2) modes during the reversal.}   
\label{fig:rev_snap}
\end{figure}

The flow structure at various stages during the reversal is shown in Fig.~\ref{fig:rev_snap}. The vorticity levels are represented with different colors and the velocity field is indicated with arrows. Fig.~\ref{fig:rev_snap}(a) shows the LSC rotating in the counterclockwise direction. The single eddy breaks into four eddies as shown in Fig.~\ref{fig:rev_snap}(b). This structure is followed transiently by two-eddy structures (Fig.~\ref{fig:rev_snap}(b-e)). Finally the upper eddy starts pushing the lower one and a single eddy rotating clockwise occupies the box (see Fig.~\ref{fig:rev_snap}(f)). Thus this suggests $[(1,1)\rightarrow(2,2)\rightarrow(1,2)\leftrightarrow(2,1)\rightarrow-(1,1)]$ to be a possible sequence of the flow structures during the reversal.

\section{Transition from the reversal regime to the condensate state}\label{condensate}

As we increase $Rh$, the flow stops reversing beyond some $Rh$, and we observe a condensate state. The condensed states are exhibited in Fig.~\ref{fig:phase_diag} using ($\blacktriangle$), and their flow profile is displayed in Figs.~\ref{fig:phase_space_lowRh}(f) and ~\ref{fig:phase_space_lowRe}(c,d).  

\begin{figure}[htbp]
\begin{center}
\includegraphics[scale=0.40]{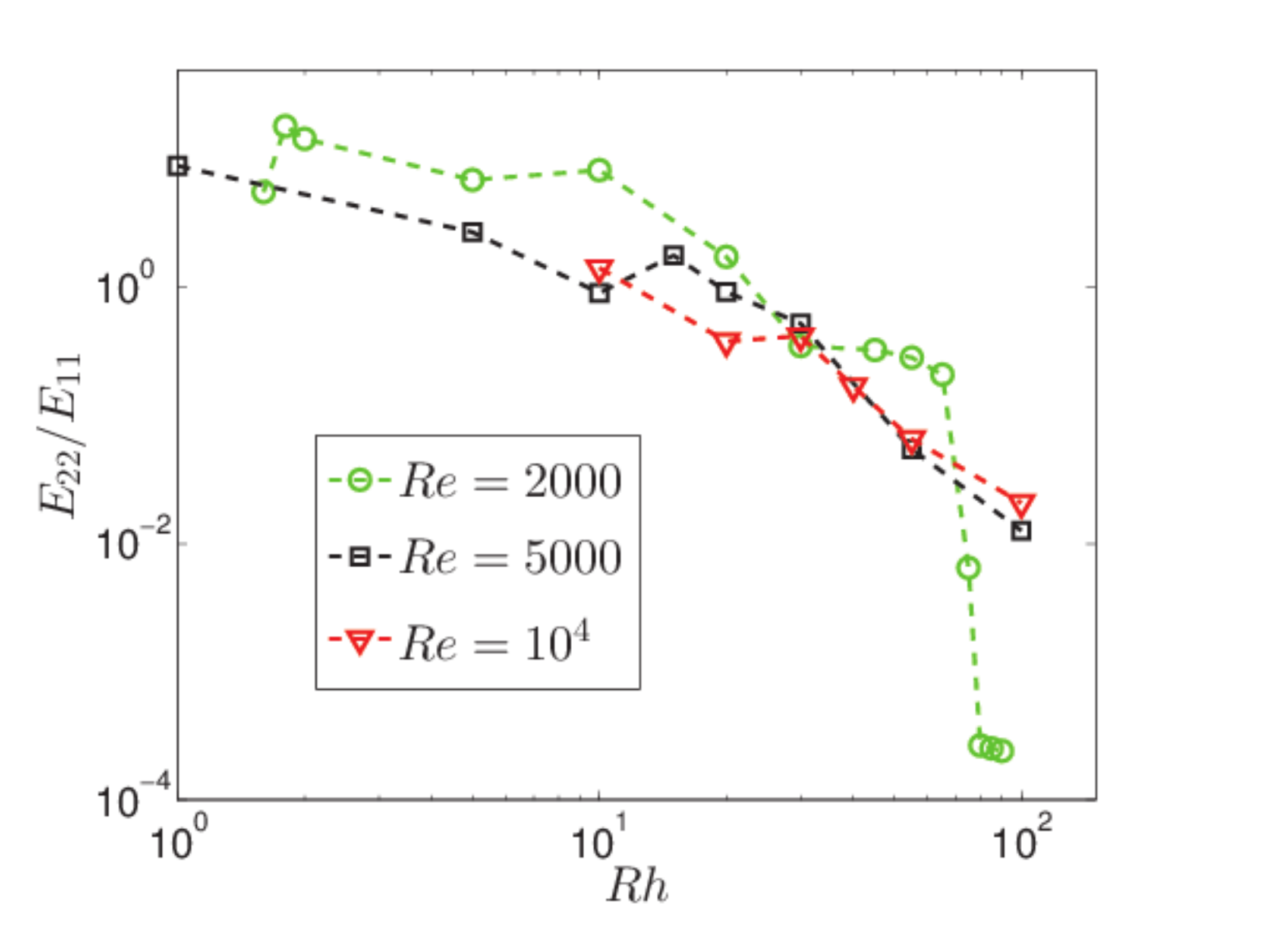}
\end{center}
\caption{Relative energy ($E_{22}$/$E_{11}$) of the mode $\hat{u}_{22}$ with respect to $\hat{u}_{11}$ versus $Rh$ for different values of $Re$. The smallest values of the relative energy are reached in the condensate state.}   
\label{fig:ratioE22byE11}
\end{figure}

As discussed in the previous section, for small $Rh$,  flow reversals occurs due to nonlinear interactions among  the modes of the triads $\{(1,1), (2,2), (3,1) \}$ and the triad $\{(1,1), (2,2), (1,3) \}$.  For the reversal, it is critical that the secondary mode $(2,2)$ be of significant strength.  We compute this ratio $|\hat{u}_{22}/\hat{u}_{11}|$ and plot it as a function of $Rh$ for $Re=2000$, $5000$ and $10^4$ in Fig.~\ref{fig:ratioE22byE11}.   We observe that the ratio $|\hat{u}_{22}/\hat{u}_{11}|$ decreases with increasing $Rh$, and it is very small  ($\sim 10^{-2}$) for large $Rh \gtrsim 100$.   This is the reason why reversals stop at large $Rh$, and we obtain a condensate state.  Note that the relative strengthening of  $\hat{u}_{11}$  mode compared to the secondary modes occurs due to an inverse cascade of energy.  Very similar phenomena was observed by Chandra and Verma~\cite{chandra:2011,chandra:2013} for thermal convection in two-dimensional box geometry; the reversal stopped at large 
Rayleigh number due to small value of the ratio $|\hat{u}_{22}/\hat{u}_{11}|$.

\begin{figure}[htbp]
\begin{center}
\includegraphics[scale=0.40]{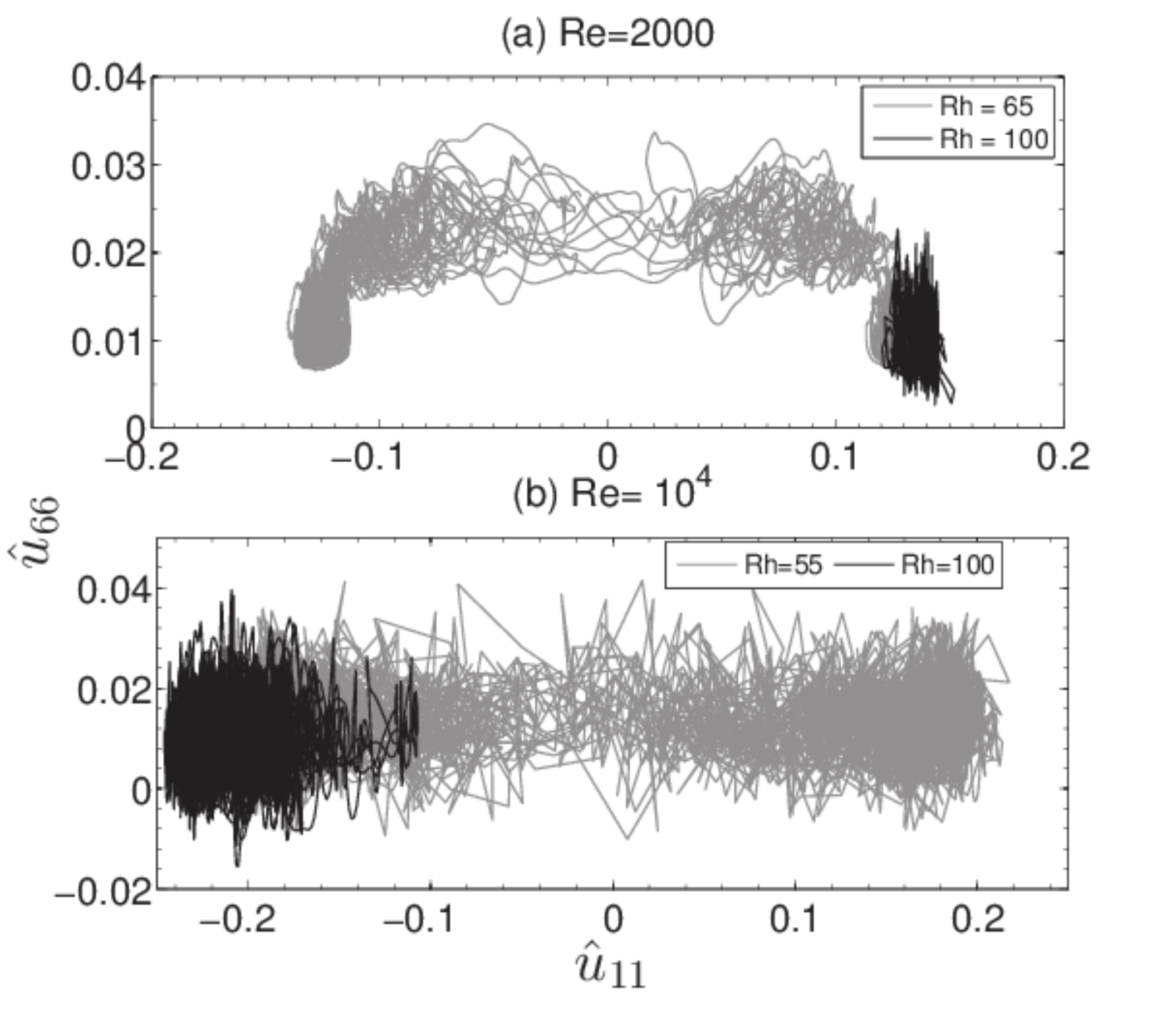}
\end{center}
\caption{Phase space projection on the modes $\hat{u}_{11}$-$\hat{u}_{66}$ for the reversal regime ($Rh=65$) and the condensate state ($Rh=100$) for $Re=2000$ (a) and $Re=10^4$ (b).} 
\label{fig:phase_space_condensate}
\end{figure}

The phase space projection on the modes $\hat{u}_{11}$-$\hat{u}_{66}$ shown in Fig.~\ref{fig:phase_space_condensate} compares the attractor related to the condensate state (in black) to the one related to the reversal regime (in grey).  The fluctuations in the condensed state increase when $Re$ is increased. For $Re=2000$, the attractor corresponding to the low energy state of the reversal regime transforms into the attractor of the condensate. 

\begin{figure}[htbp]
\begin{center}
\includegraphics[scale=0.40]{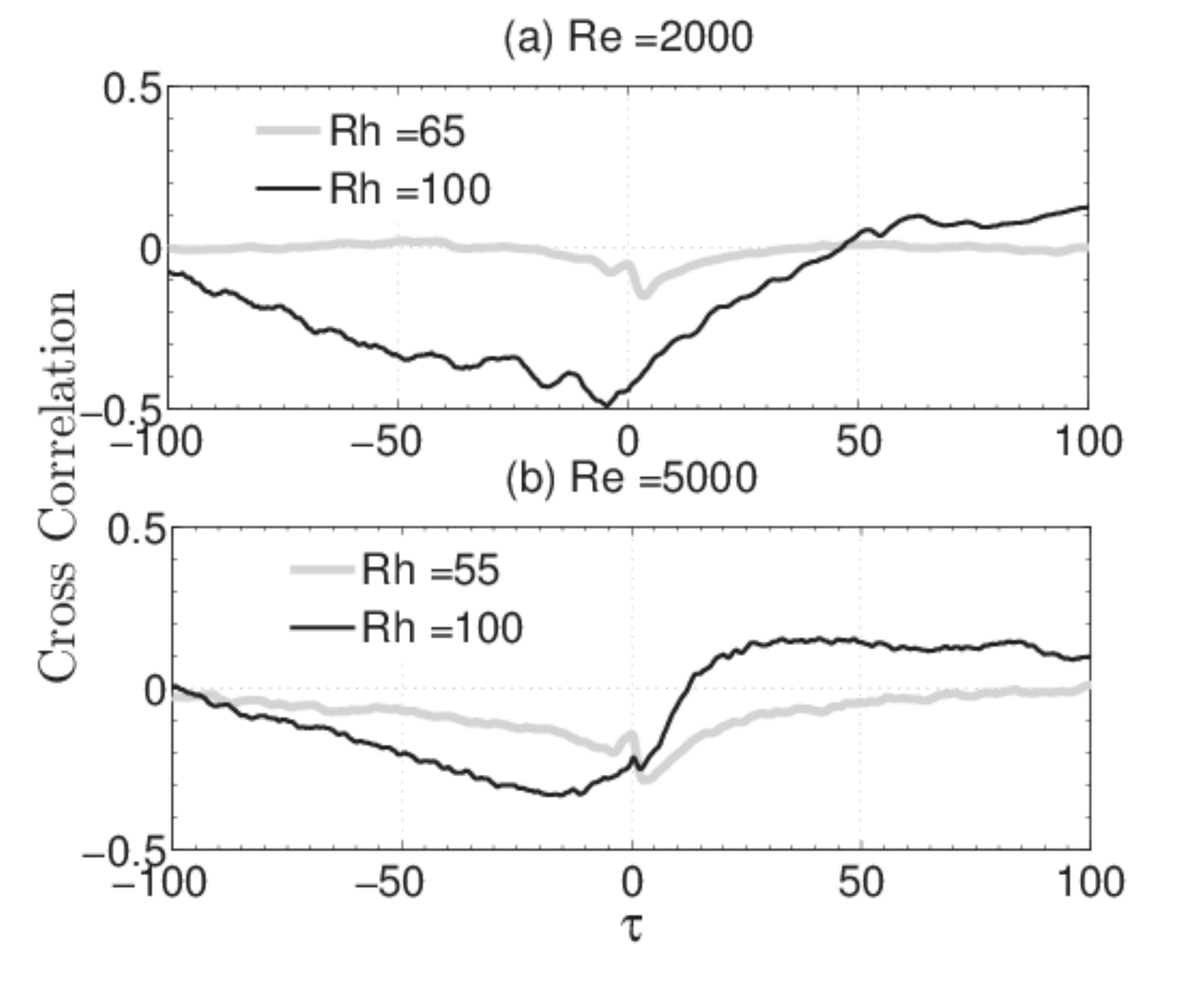}
\end{center}
\caption{Cross-correlation between the injected power $I$ and the large scale kinetic energy $\hat{u}_{11}^2$ in the reversal regime (grey) and in the condensate state (black) for $Re=2000$ (a) and $Re=5000$ (b).}
\label{fig:correlationrevcond}
\end{figure}

Finally, it is instructive to consider the cross-correlation between the injected power $I$ and the kinetic energy of the large scale mode $\hat{u}_{11}^2$ in the reversal regime and the condensate state. Fig.~\ref{fig:correlationrevcond} shows that the large scale flow and the injected power are anticorrelated, i.e. the cross-correlation reaches large negative values for some time lag. This is in agreement with the mechanism proposed by Tsang and Young~\cite{tsang:2009} in which the large scale flow detunes the small scale pattern from the forcing thus decreasing the injected power. However, the sign of the time lag changes with $Rh$. In the reversal regime, it is positive, thus the kinetic energy of the large scale mode lags behind the injected power. Consequently, reversals may result from fluctuations in the energy budget.  The opposite is observed 
in the condensate state for which the injected power lags behind the large scale flow. Detuning induced by the large scale flow controls the injected power.

As described above, the condensate state has been predicted in the framework of two-dimensional turbulence phenomenology~\cite{kraichnan:1967}.
Its observation in direct numerical simulations has been first performed using the two-dimensional Navier-Stokes equation without large scale drag and using hyperviscosity to regularize the small scales~\cite{smith:1993, chertkov:2007}. In addition, the forcing was applied at small scale with respect to the box size and consists of white noise. The later provides a constant injected power and scale separation and hyperviscosity allow a well developed inverse cascade of energy. We have shown here that a condensate can be obtained even when both the large scale drag and small scale viscous dissipation play a significant role in the energy budget. A different approach for the condensed state has been recently provided by showing that it can be understood has a nonlinear solution of the two-dimensional Navier-Stokes equation without large scale drag. In the limit $Re \rightarrow \infty$, the LSC detunes the linear response to the constant and spatially periodic forcing, such that the injected power becomes 
vanishingly 
small~\cite{gallet:2013}. This approach strongly differs from the phenomenology of two-dimensional turbulence as far as the injected power is concerned. It looks to fit better with the flow configuration we studied in this work.

\section{Conclusions} 

We performed a numerical study of a two-dimensional Kolmogorov flow generated by a spatially periodic and constant force and linearly damped by a drag force with a decay rate $1/Rh$. When the Reynolds number $Re$ or $Rh$ are small, the linear response to the forcing is stable, and the system takes the form of a square arrays of $6\times6$ alternate counter-rotating vortices.  We studied variation of patterns for various $Re$ and $Rh$, but focussed on varying $Rh$ for a particular $Re$, more specifically $Re=5000$. 
When $Rh$ was increased, the flow becomes unstable, and it follows a sequence of  transitions to a fixed point, periodic, quasiperiodic, and chaotic states.  The most dominant Fourier mode $\hat{u}_{11}$ that represents the large-scale flow describes the aforementioned states very well.

We observed two kinds of chaotic states: a first one with a nonzero mean value of $\hat{u}_{11}$, and the second one with a zero mean value of $\hat{u}_{11}$.  The second chaotic state occurs for larger $Rh$. On further increase of $Rh$, the PDF of the mode $\hat{u}_{11}$ transitions from a gaussian distribution to a bimodal distribution.   The maxima of the bimodal distribution are related to LSC with opposite values of the velocity.  The distribution becomes more and more peaked as $Rh$ is increased, thus leading to flow reversals, in which the mode $\hat{u}_{11}$ randomly switches between two quasi-stationary states.

A further increase in $Rh$ to around 100 leads to cessation of flow reversals, and we obtain a condensate state in which the flow moves continuously in one direction.  We attribute the condensate state to weakening of the secondary mode $\hat{u}_{22}$ that plays a very critical role in the flow reversal.  The relative strengthening of $\hat{u}_{11}$ over the secondary modes is due to the inverse cascade of energy in two-dimensional flow. 

 Very similar phenomena were observed by Chandra and Verma~\cite{chandra:2011,chandra:2013} for turbulent convection in a two-dimensional square box. They obtained flow reversals after a series of bifurcations, as the Rayleigh number was increased.  However, the flow reversals stopped for very large Rayleigh number due to weakening of the secondary mode $\hat{u}_{23}$.  Chandra and Verma~\cite{chandra:2011,chandra:2013} argue that the most dominant triads responsible for the flow reversals in turbulent convection are ${(1,1), (2,2), (1,3)}$ and ${(1,1), (2,2), (3,1)}$, which appears to be the same in the Kolmogorov flow discussed in this paper.

All these transitions are in fair agreement with experimental observations~\cite{sommeria:1986,herault:2013} thus showing that the two-dimensional modeling of the flow with stress-free boundary conditions is a reasonable approximation. Taking into account rigid boundary conditions can be important in the absence of large scale drag ($1/Rh = 0$) when reversals are generated by random forcing~\cite{molenaar:2004,vanheijst:2006}. In that case, it has been found that reversals are induced by generation of vorticity in the lateral boundary layers. However, this does not correspond to the experimental configurations quoted above or to our numerical simulations. 

Reversals of a vector field are widely observed in geophysics or astrophysics. The so-called quasi-biennial oscillation is an example of a large scale
almost cyclic reversing flow in the turbulent atmosphere. It is a roughly periodic oscillation in the strength and direction of the zonal (east-west) wind in the lower and middle stratosphere over the equator of the Earth's atmosphere~\cite{baldwin:2001}. The magnetic field of the Earth and the sun also display random (respectively roughly periodic) reversals. Reversals of a large scale velocity field over a turbulent background in thermal convection~\cite{sugiyama:2010,chandra:2011,chandra:2013} or of a magnetic field generated through the dynamo effect by a von Karman swirling flow~\cite{berhanu:2010} have been studied in the past years in laboratory experiments. Although the dynamo experiment involves flows with much larger turbulent fluctuations than the convection experiment or the Kolmogorov flow, it has been found that reversals of the magnetic field follow well defined trajectories in phase space (see Fig. 7 in ~\cite{berhanu:2010}). This clearly differs from the reversals studied here (see Fig.~\ref{fig:phase_space_2e03}). 
Although we have identified some intermediate structures that correspond to the Fourier, modes (2,1), (1,2) and (2,2) during reversals, it seems unlikely that low dimensional models similar to the ones designed for reversals of the magnetic field~\cite{petrelis:2008,petrelis:2010} can capture all the features of reversals of the LSC in Kolmogorov flows. Turbulent fluctuations are difficult to disentangle from the dynamics of the dominant modes in the present situation, such that a different approach should be found to describe these bifurcations between different turbulent regimes.

\appendix
\section{Energy transfers among Fourier modes in Kolmogorov flows}

As described in Sec.~\ref{numdetails}, we force the $(6,6)$ Fourier mode, hence the energy input to the system is via this mode.  The nonlinearity however enables energy transfers to other modes.  In this paper we focus on the energy transfer to the (1,1) mode, which contributes most significantly to the large-scale circulation (LSC).

The equation for the energy of a mode $\mathbf k = (k_x,k_y)$ is
 \begin{equation}
  \frac{\partial}{\partial t} \frac{1}{2} {\hat{\mathbf u}(\mathbf k)}^2 = T(\mathbf k) - \left( \frac{1}{Rh} + \frac{k^2}{Re} \right) {\hat{\mathbf u}(\mathbf k)}^2 + \hat{\mathbf f}(\mathbf k) \cdot \hat{\mathbf u}(\mathbf k) ,\label{eq:energy_u}
 \end{equation}
where $\hat{\mathbf u }(\mathbf{k}) = (\hat{u}(k),\hat{v}(k))$ are the Fourier component of the velocity, $\hat{\mathbf f}(\mathbf{k})$ is the Fourier component of the force, and $T(\mathbf k)$ is the rate of energy transfer into $\hat{u}(\mathbf k)$ mode from all other modes via nonlinear triad interactions.  Dar et al. \cite{dar:2001} and Verma \cite{verma:2004} showed that 
 \begin{equation}
T(\mathbf k)  = \sum_{\mathbf p} S(\mathbf k| \mathbf p| \mathbf q) 
 \end{equation}
 with $\mathbf q = \mathbf k - \mathbf p$, and
 \begin{equation}
S(\mathbf k| \mathbf p| \mathbf q)  = -[(\mathbf{k}\cdot\mathbf{\hat{u}(\mathbf q)})(\mathbf{\hat{u}(\mathbf k)}\cdot\mathbf{\hat{u}(\mathbf p)}]\label{eq:transfer_uk}
 \end{equation} 
 is the ``mode-to-mode energy transfer" from mode $\mathbf p$ to $\mathbf k$ with $\mathbf q$ acting as a mediator.   Note that $\hat{\mathbf u} (\mathbf k) =( \hat{u}(\mathbf k), \hat{v}(\mathbf k)) $ is a  real vector.  Using the incompressibility condition [Eq.~(\ref{eq:NS_incom})], the above expression becomes
\begin{equation}
  S(\mathbf{k}|\mathbf{p}|\mathbf{q})=\pi k_x \left( \frac{k_y q_x}{ k_x q_y}- 1 \right) \left( \frac{p_x k_x}{p_y k_y} + 1\right) {\hat{u}(\mathbf q)}{\hat{u}(\mathbf k)}{\hat{u}(\mathbf p)}.  \label{eq:transfer_u11_v1}
 \end{equation}
 
The above expression for the mode-to-mode energy transfer is very useful.  We can deduce the energy transfer rate from the $(6,6)$ mode to the $(1,1)$ mode using this formula.  Clearly, $\{ (1,1), (6,6), (5,7) \}$ and $\{ (1,1), (6,6),  (7,5) \}$ form two important triads for the above energy transfers.  Hence,  
 \begin{eqnarray}
S( (1,1)| (6,6) | (5,7))  & = & -\frac{4 \pi}{7}  \hat{u}(1,1)}{\hat{u}(6,6)}{\hat{u}(5,7), \label{eq:66to11}\\
S( (1,1)| (5,7) | (6,6))  & = & 0. \label{eq:57to11} \\
S( (6,6)| (1,1) | (5,7))  & = & - S( (1,1)| (6,6) | (5,7)) \\
S( (1,1)| (6,6) | (7,5))  & = & \frac{4 \pi}{5}  \hat{u}(1,1)}{\hat{u}(6,6)}{\hat{u}(7,5), \\
S( (1,1)| (7,5) | (6,6))  & = & 0. \\
S( (6,6)| (1,1) | (7,5))  & = & -S( (1,1)| (6,6) | (7,5)) 
 \end{eqnarray} 
 It is interesting to note that in the triad $\{ (1,1),(6,6),(5,7) \}$, $(6,6)$ mode supplies energy to $(1,1)$ [Eq.~(\ref{eq:66to11})], but $(5,7)$ does not supply energy to $(1,1)$ [Eq.~(\ref{eq:57to11})].  
   
We have not analysed the energy exchanges among all the modes in great detail.  However our analysis shows that $(1,1)$ is the most important mode and it receives energy from the $(6,6)$ mode.  Under steady state, the energy received by $(1,1)$ mode is transferred to higher modes, e.g. $(2,2)$.  These modes in turn transfer energy to other modes.

\section{Symmetry of flow reversals in Kolmogorov flow}
The dynamical equation in Fourier space corresponding to Eq.~(1) is given by 
 \begin{eqnarray}
  \frac{\partial \hat{u}_j(\mathbf k)}{\partial t} & = & - k_i \sum_{\mathbf{p+q=k}} \hat{u}_j(\mathbf q) \hat{u}_i(\mathbf p) - k_i \sigma(\mathbf k) \\ \nonumber
&&  - \left( \frac{1}{Rh} + \frac{k^2}{Re} \right) \hat{u}_i(\mathbf k) + \hat{f}_i(\mathbf k) 
 \end{eqnarray}
Symmetry arguments can be very useful in deducing the nature of Fourier modes after the reversal, specially when we apply constant forcing to a single Fourier mode (e.g., $\mathbf f(\mathbf k_0)$), or to a small number of modes.   These arguments are based on geometrical symmetries like reflections or rotation~\cite{gallet:2012} , or ``even-odd'' symmetries of the interacting Fourier modes~\cite{verma:2014}. In the following discussion, we generalise the arguments of Verma et al.~\cite{verma:2014} for the Kolmogorov's flow.

For the 2D box, the Fourier modes come in four categories: $\{ E \}=$ (even, even), $\{ O \}=$ (odd, odd), $\{ M_{1}  \}= $ (even, odd), and $\{ M_{2} \}= $ (odd, even). For example, (1,1) is an odd mode, and (2,2) is an even mode. These  elements form an abelian group called ``Klein four-group", which is a direct product of two cyclic groups of  two elements each, i.e. $Z_2 \times Z_2$ (see Table 1).  
%

\begin{table}
\caption{Rules of nonlinear interactions among the Fourier modes in 2D Kolmogorov flow. The elements form the Klein four-group $Z_2 \times Z_2$.} 
\label{tab:2d_product_rule}
\begin{tabular}{|a | c | c | c | c|}
\hline
\rowcolor{Gray}
$\times$ & $E$ & $M_1$ & $M_2$ & $O$ \\ \hline
$E$ & $E$ & $M_1$ & $M_2$ & $O$ \\ \hline
$M_1$ & $M_1$ & $E$ & $O$ & $M_2$ \\ \hline
$M_2$ & $M_2$ & $O$ & $E$ & $M_1$ \\ \hline
$O$ & $O$ & $M_2$ & $M_1$ &  $E$ \\ \hline
\end{tabular}
\end{table}

When a particular Fourier mode is forced with a constant forcing, i.e. $\hat{f}_i(\mathbf k)$, then the reversal rules depend on which class the forcing wavenumber belong to.  Using the product rule, we can make the following deductions on the reversing modes.

\begin{enumerate}

\item $\{ E \}$  modes cannot change sign during a reversal.  

\item If a constant forcing is applied to $\{ E \}$  modes (e.g., $\mathbf f(6,6) =$ const.), then the Fourier modes can belong to one of the six classes: 

$ \{ O \} \rightarrow  \{ -O \}$; $ \{ E \} \rightarrow  \{ E\}$; $ \{ M_1, M_2 \} =\epsilon$; 

$ \{ M_1 \} \rightarrow  \{ -M_1 \}$; $ \{ E \} \rightarrow  \{ E\}$; $ \{ O, M_2 \} =\epsilon$; 

$ \{ M_2 \} \rightarrow  \{ -M_2 \}$; $ \{ E \} \rightarrow  \{ E\}$; $ \{ O, M_1 \} =\epsilon$; 

$ \{ O \} \rightarrow  \{ -O \}$; $ \{ M_1 \} \rightarrow  \{ -M_1 \}$; $ \{ M_2 \} \rightarrow  \{ M_2\}$; $ \{ E \} \rightarrow  \{ E\}$; 

$ \{ O \} \rightarrow  \{ -O \}$; $ \{ M_2 \} \rightarrow  \{ -M_2 \}$; $ \{ M_1 \} \rightarrow  \{ M_1\}$; $ \{ E \} \rightarrow  \{ E\}$;  

$ \{ M_1 \} \rightarrow  \{ -M_1 \}$; $ \{ M_2 \} \rightarrow  \{ -M_2 \}$; $ \{ O \} \rightarrow  \{ O\}$; $ \{ E \} \rightarrow  \{ E\}$,

where $\epsilon$ is a small number, and it represents fluctuating modes around zero.

The above conclusion also follows  from the fact that $\{ M_1 \}$ and $\{ M_2 \}$  are symmetric under mirror reflections $S_x$ and $S_y$ respectively, while the $\{ O \}$ modes are odd under $S_x$ and $S_y$ symmetry operations~\cite{gallet:2012}. 

\item If a constant forcing is applied to $O$ type mode, then  $\{ O \}$  modes do not change sign.  Clearly,  $\{ E \}$  modes do not change sign.  Hence, $ \{ M_1 \} \rightarrow  \{ -M_1 \}$; $ \{ M_2 \} \rightarrow  \{ -M_2 \}$; $ \{ O \} \rightarrow  \{ O\}$; $ \{ E \} \rightarrow  \{ E\}$.

Hence,  $\{ M_1 \}$ and $\{ M_2 \}$ change during a reversal.  It follows from the fact that  $\{ O \}$ modes are even under rotation by $\pi$, but  $\{ M_1 \}$ and $\{ M_2 \}$ are odd under this operation.

\item For a constant forcing of the  $\{ M_1 \}$ type mode, then $ \{ O \} \rightarrow  \{ -O \}$; $ \{ M_2 \} \rightarrow  \{ -M_2 \}$; $ \{ M_1 \} \rightarrow  \{ M_1\}$; $ \{ E \} \rightarrow  \{ E\}$. 
 Similarly, for constant forcing of  $\{ M_2 \}$ type modes,  $ \{ O \} \rightarrow  \{ -O \}$; $ \{ M_1 \} \rightarrow  \{ -M_1 \}$; $ \{ M_2 \} \rightarrow  \{ M_2\}$; $ \{ E \} \rightarrow  \{ E\}$.
 
\item For mixed constant forcing like $\mathbf f(E) + \mathbf f(O)$, the modes follow the reversal rules same as that for constant forcing of odd modes.  However, the reversals would be suppressed under forcing of the type $\mathbf f(O) + \mathbf f(M_{1})$ or $\mathbf f(O)+\mathbf f(M_{2})$.
\end{enumerate}

The above rules cover all the symmetry rules for constant forcing.  The present paper deals with $\mathbf f(6,6)$ forcing which comes under  $\{ E \}$ category.  Our result shows that the $(1,1)$ mode reverses,  the $(2,2)$ mode maintains the sign of its average value, while $(1,2)$ and $(2,1)$  fluctuate around zero during reversals. Hence, the Fourier modes under the forcing of $f(6,6)$ mode belong to $ \{ O \} \rightarrow  \{ -O \}$; $ \{ E \} \rightarrow  \{ E\}$; $ \{ M_1, M_2 \} =\epsilon$ class. 
\bigskip

\textbf{Acknowledgments:} 
Support of IFCPAR/CEFIPRA contract 4904-A is acknowledged. MKV thanks ENS Paris for the kind hospitality provided during his visit when this work was completed, and thank IFCPAR/CEFIPRA for the travel support to ENS Paris.

\end{document}